\documentclass[acmtog, nonacm]{acmart}

\makeatletter
\renewcommand\@formatdoi[1]{\ignorespaces}
\makeatother

\setcitestyle{square}

\usepackage{acmart-taps}

\usepackage[utf8]{inputenc}
\usepackage{algorithm}
\usepackage{algorithmicx}
\usepackage{algpseudocode}

\usepackage{natbib}
\usepackage{subcaption}
\usepackage{afterpage}
\usepackage{xcolor,colortbl}
\usepackage{ifthen}
\usepackage{siunitx}
\usepackage[capitalize,noabbrev]{cleveref}

\crefname{requirement}{Requirement}{Requirements}
\Crefname{requirement}{Requirement}{Requirements}

\usepackage{multirow}
\usepackage{arydshln} %
\usepackage{makecell} %
\usepackage{pgfplots}
\pgfplotsset{compat=1.18}
\usepackage{wrapfig}
\usepackage{xspace}

\usepackage{xr} %

\definecolor{ZizhouColor}{rgb}{0.15, 0.68, 0.38} 
\definecolor{ZachColor}{rgb}{0.9, 0.5, 0}
\definecolor{MaxColor}{rgb}{0, 0, 0.8}
\definecolor{DenisColor}{rgb}{0.15, 0.38, 0.68}
\definecolor{DanieleColor}{rgb}{1, 0, 0}

\definecolor{MATLABColor1}{HTML}{0072BD}
\definecolor{MATLABColor2}{HTML}{D95319}
\definecolor{MATLABColor3}{HTML}{EDB120}
\definecolor{MATLABColor4}{HTML}{7E2F8E}
\definecolor{MATLABColor5}{HTML}{4DBEEE}
\definecolor{MATLABColor6}{HTML}{A2142F}

\newcommand{\DP}[1]{}
\newcommand{\DZ}[1]{}
\newcommand{\ZH}[1]{}
\newcommand{\MP}[1]{}
\newcommand{\ZF}[1]{}

\newcommand{\he}{h_\epsilon}
\newcommand{\barrier}{p}
\newcommand{\pex}{\barrier_{\epsilon(x)}}
\newcommand{\pe}{\barrier_{\epsilon(x)}}

\newcommand{\pipc}{\barrier^{\text{IPC}}_\epsilon}
\renewcommand{\paragraph}[1]{\vskip0.5em\noindent\textbf{#1.}}

\aptLtoXcmd{}{%
}

\aptLtoXcmd{}{%
}

\aptLtoXcmd{}{}

\let\vec\mathbf
\newcommand{\transposeSymbol}{\mathstrut\mathbf{\top}}
\newcommand{\transpose}[1]{#1^{\transposeSymbol}}

\usepackage{acronym}
\newacro{IP}{incremental potential}
\newacro{IPC}{Incremental Potential Contact}
\newacro{CIPC}[C-IPC]{Codimensional \ac{IPC}}
\newacro{NURBS}{non-uniform rational B-spline}
\newacro{IGA}{isogeometric analysis}
\newacro{DOF}{degrees of freedom}
\newacro{FE}{finite element}
\newacro{FEM}{finite element method}
\newacro{CCD}{continuous collision detection}
\newacro{LCP}{linear complementarity problem}
\newacro{LBFGS}[L-BFGS]{Limited-memory BFGS}
\newacro{HPC}{high performance computing}
\newacro{HO}{high-order}

\newcommand{\figname}[1]{\textbf{#1.}}

\newcommand{\paperTitle}{Geometric Contact Potential}

\theoremstyle{definition}
\newtheorem{definition}{Definition}

\theoremstyle{remark}
\newtheorem*{remark}{Remark}

\theoremstyle{definition}
\newtheorem{requirement}{Requirement}

\theoremstyle{theorem}
\newtheorem{prop}{Proposition}

\graphicspath{{./figs/}{./figs/sketches}}

\newcommand{\dhat}{\ensuremath{\hat{d}}}

\newcommand{\Rgen}{\emph{Finiteness}\xspace}
\newcommand{\Rbar}{\emph{Barrier}\xspace}
\newcommand{\Rrest}{\emph{No spurious forces}\xspace}
\newcommand{\Rloc}{\emph{Localization}\xspace}
\newcommand{\Rdiff}{\emph{Differentiability}\xspace}
\newcommand{\Rindep}{\emph{Resolution independence}\xspace}
\newcommand{\Rdiscr}{\emph{Discretization}\xspace}

\newcommand{\domain}{\ensuremath{\operatorname{Int}(\Omega)}}
\newcommand{\boundary}{\ensuremath{\Omega}}

\newcommand{\potential}{\ensuremath{\psi_\epsilon}}

\newcommand{\interaction}{interaction}
\newcommand{\Interaction}{Interaction}

\newcommand{\smoothingfactor}{directional factor}
\newcommand{\smoothingfactors}{directional factors}
\newcommand{\Smoothingfactor}{Directional factor}



\title{\paperTitle}

\author{Zizhou Huang}
\email{zizhou@nyu.edu}
\affiliation{%
  \institution{New York University}
  \country{USA}
}

\author{Maxwell Paik}
\email{mp6569@nyu.edu}
\affiliation{%
  \institution{New York University}
  \country{USA}
}

\author{Zachary Ferguson}
\email{zy.fergus@gmail.com}
\affiliation{%
  \institution{Massachusetts Institute of Technology}
  \country{USA}
}

\author{Daniele Panozzo}
\email{panozzo@nyu.edu}
\affiliation{%
  \institution{New York University}
  \country{USA}
}

\author{Denis Zorin}
\email{dzorin@cs.nyu.edu}
\affiliation{%
  \institution{New York University}
  \country{USA}
}

\renewcommand\footnotetextcopyrightpermission[1]{} 
\begin{document}

\begin{abstract}

Barrier potentials gained popularity as a means for robust contact handling in physical modeling and for modeling self-avoiding shapes. 
The key to the success of these approaches is adherence to geometric constraints, i.e., avoiding intersections, which are the cause of most robustness problems in complex deformation simulation with contact. However, existing barrier-potential methods may lead to spurious forces and imperfect satisfaction of the geometric constraints. They may have strong resolution dependence, requiring careful adaptation of the potential parameters to the object discretizations. 

 We present a systematic derivation of a continuum potential defined for smooth and piecewise smooth surfaces, starting from identifying a set of natural requirements for contact potentials, including the barrier property, locality, differentiable dependence on shape, and absence of forces in rest configurations. Our potential is formulated independently of surface discretization and addresses the shortcomings of existing potential-based methods while retaining their advantages. 

We present a discretization of our potential that is a drop-in replacement for the potential used in the incremental potential contact formulation \cite{Li2020IPC}, and compare its behavior to other potential formulations, demonstrating that it has the expected behavior.  The presented formulation connects existing barrier approaches, as all recent existing methods can be viewed as a variation of the presented potential, and lays a foundation for developing alternative (e.g., higher-order) versions. 
\end{abstract}

\maketitle

\thispagestyle{empty}

\acresetall %

\section{Introduction}
\label{sec:intro}
\begin{figure}
    \centering
    \includegraphics[width=\linewidth]{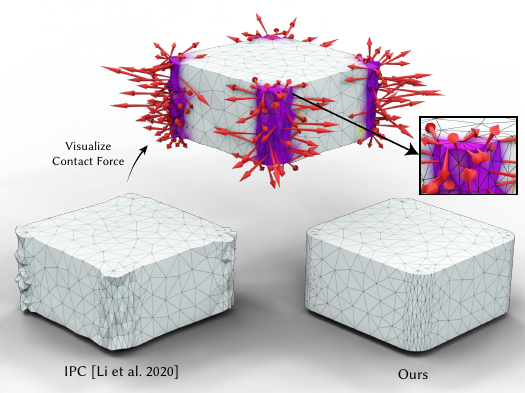}
    \caption{
    We introduce a novel geometric barrier potential satisfying a set of natural properties. None of the other potentials in the literature satisfy all these properties simultaneously, leading to inaccurate results, penetrations, or other undesired artifacts. In this example, we show that the contact potential introduced in \cite{Li2020IPC} is not zero at the rest pose when the potential extent is larger than the length of an edge, introducing spurious forces (top) and deformation (bottom left). Our potential is zero in the rest pose by construction (bottom right), as this is one of the natural properties of a geometric contact potential.
    }
    \label{fig:teaser}
\end{figure}

Contact modeling is a critical component of simulation tools for many domains, including computer graphics, robotics,  mechanical design, and biomedical engineering.  A representative (friction-less) contact problem can be viewed as a geometrically constrained optimization:
\[ \min_u E(u),\; \mbox{subject to $g(u) \geq 0$}
\]
where $u = u(x)$ is a deformation of a surface parametrized by $x$, and the function $g$ is entirely geometric, measuring the \emph{distance to contact} for a solution $u$.
 $E(u)$ may be physical energy for static problems, a time integrator formulated in a variational way for dynamic problems, or a deformation objective for geometric modeling.
 
 Overwhelmingly, robustness issues in contact modeling are related to handling these geometric constraints; once a contact constraint is violated, it may be very difficult for the solver to recover.  For this reason, a natural approach to increasing the robustness of contact solvers is to ensure that \emph{all solution updates}, including intermediate steps of iterations in a solver, are \emph{contact-free}, i.e., satisfy $g(x) > 0$. 

Solvers based on barrier potentials are particularly suitable to maintain a contact-free state, and in recent years, were demonstrated to handle complex, large-scale, and long-duration contact problems reliably. Conceptually, these solvers are based on viewing the contact problem as an equivalent unconstrained optimization \cite{Kane1999Finite}: 
\[ \min_u E(u) + b(u)
\]
where $b(u)$ is an ideal barrier, equal to $\infty$ for configurations with interpenetration, and $0$ for contact-free. 

If we take $b(u)$ to be a potential that increases smoothly to infinity as $u$ approaches a contact configuration, then the problem is converted to a smooth problem that can be solved with contact-free iterates, e.g., using a nonlinear solver with line search, equipped with continuous collision detection (CCD), as it is done in \cite{Li2020IPC}. 

All existing contact barrier potentials are defined based on aggregating repulsion terms between pairs of points or, in the discrete case, between pairs of elements (e.g., a face and a point).  The key question then becomes how to choose the strength of repulsion between two points so that it is high when these points are close to contact, vanishes when these points are far, and has a number of other desirable properties (e.g., depends smoothly on the deformation, has controlled locality), as discussed in \cref{sec:requirements}. 

\begin{wrapfigure}{R}{0.15\textwidth}
    \includegraphics[clip,width=0.15\textwidth]{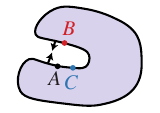}
    \caption{Contact potentials must distinguish points nearing contact (A and B) from nearby non-contacting points (A and C).}
    \label{fig:nearby-points}
\end{wrapfigure}

One important difficulty is distinguishing true contact, arising from different objects or parts of the surface of the same object moving toward each other, and points that remain close because they are close on the undeformed rest shape (\cref{fig:nearby-points}). Answering this question for a general class of surfaces requires careful geometric consideration.  In Section~\ref{sec:related-barrier}, we review how it is addressed for existing potential types and problems associated with these approaches (possibility of geometric constraint violations, spurious forces, and strongly discretization-dependent behavior).

We undertake a systematic derivation of a \emph{continuum barrier potential} defined as a surface integral,  and its discretization, which retains key features of previously proposed approaches, IPC in particular, but satisfies several additional, natural requirements, elaborated in \cref{sec:contact-candidates}:
\begin{itemize} 
    \item \Rindep:  The potential definition is independent of discretization. 
    \item \Rgen: The total barrier potential over the surface is defined and finite for any collection of piecewise-smooth surfaces not in contact. (The use of piecewise-smooth, rather than smooth, surfaces is critical for many applications, as detailed in \cref{sec:pw-sm}). 
    \item \Rbar: The potential grows to infinity as the objects in the simulation approach (self-)contact,  with respect to a suitably defined \emph{distance-to-contact}. 
    \item \Rloc: The potential has a localization parameter $\hat{d}$, which may vary over the object surface. The potential vanishes if the objects are further away than $\hat{d}$ from contact, with respect to the same measure of contact.
    \item \Rrest: In the undeformed configuration, the potential is zero.
    \item \Rdiff: The potential depends smoothly on the surface configuration (e.g., for piecewise-linear surfaces, on mesh vertex positions).
\end{itemize}

The foundation of our approach is a \emph{new geometric definition of the distance-to-contact} that allows us to define contact potentials for arbitrary piecewise-smooth surfaces without assuming a discretization, and in a way that all requirements are satisfied, leading to artifact-free simulation with a flexible choice of potential locality.  The key features of our approach include:

\begin{itemize} 
\item  A systematic way of defining points close to contact for smooth and piecewise-smooth surfaces, based on \emph{interaction sets}.
\item  An approach for defining a contact potential for piecewise-smooth surfaces that meets the requirements enumerated above. 
\item A discrete version of this potential for piecewise-linear surfaces that satisfies the requirements enumerated above and has similar efficiency to standard IPC.
\end{itemize}

We demonstrate that the new formulation eliminates spurious forces inherent in other formulations, converges to a limit under refinement, and decouples the barrier extent from the discretization choice, offering needed flexibility for applications such as shape optimization.

\section{Barrier potentials}
\label{sec:related-barrier}

To motivate the proposed approach, we consider several representative versions of contact potentials proposed in the past. 
We focus on four problems that each construction addresses differently: 
\begin{itemize}
\item \emph{Finiteness.} How the potentials ensure that true contact points are distinguished from close material points; without this distinction, they would always be infinite.  
\item \emph{Spurious repulsion.} Relatedly, how spurious forces for close material points are eliminated or reduced; if applied to the wrong pairs of points, these forces have a large impact. 
\item \emph{Mesh dependence.} How these potentials behave under remeshing or refinement; mesh adaptation, refinement, or remeshing are common, especially for large/complex deformations. 
\item \emph{Discretization properties.} If a potential is defined in continuous form, even if it prevents contact exactly, its discretization may do so only approximately. 
\end{itemize}

\subsection{The IPC barrier potential \cite{Li2020IPC}}  This potential is defined in a purely discrete way, based on vertex-face and edge-edge repulsion.  To ensure it is \emph{finite}, interactions of a face with its vertices and between edges sharing one vertex are excluded. Note that as the surface is refined, the strength of interaction increases, as the excluded part gets smaller.   

\begin{figure}
    \includegraphics[width=\linewidth]{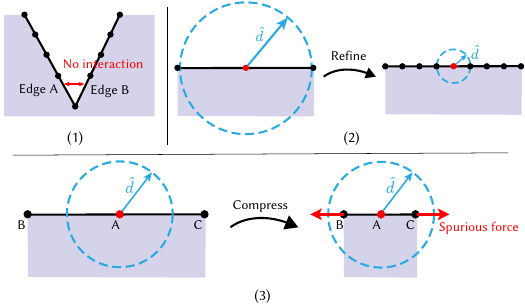}
    \caption{IPC potential.  The extent of the potential is shown in blue. 
    (1) IPC potential is finite, as interactions between edges sharing vertices, and faces with their vertices are dropped. (2) Refinement forces the maximal potential extent $\hat{d}$ to decrease. (3) Spurious forces (red arrows) arise if the surface is compressed horizontally and nearby vertices are closer than the potential extent $\hat{d}$. 
    }
    \label{fig:IPC-problems}
\end{figure}
    
\emph{Spurious forces} are partly avoided by defining a potential that does not extend beyond the shortest mesh edge length in the undeformed state.   However, during deformation, the edge length can change arbitrarily, and spurious forces may easily appear under compression (\cref{sec:spurious-eval}).  
The IPC potential is strongly \emph{mesh dependent}: e.g., simply refining locally forces the maximum extent of the potential to be decreased by a large factor, resulting in a change in the results of simulation/deformation. 

A modified formulation of this potential is introduced in \citet{Li2023Convergent}, with a convergent discretization which, however, requires refinement in both spatial discretization and the potential extent $\dhat$. We provide further discussion and comparisons in \cref{sec:eval} and \cref{tab:compare_other}.

\subsection{Surface/volume double-integral potentials \cite{sauer2013computational,Kamensky2018Contact}} 
\begin{figure}
    \begin{center}
\includegraphics[width=\linewidth]{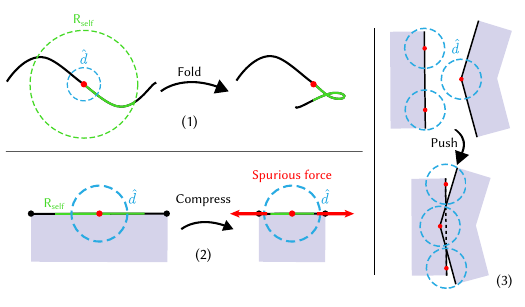}
    \end{center}    
    \caption{Surface barrier of \cite{Kamensky2018Contact} (1) The potential has a self-contact exclusion zone (green), outside the potential extent (blue). For extreme deformations (folding), this allows self-intersections. (2) Extreme compression may push points into the potential zone, leading to spurious forces (red arrows) (3) In quadrature point-to-point discretization, the surface may not remain contact-free, a point may "push through" between other \protect\cite{Kamensky2018Contact}.
    }
    \label{fig:SL-problems}
\end{figure}

Barrier potentials defined in continuum setting in
\cite{sauer2013computational} are adapted to self-contact in \cite{Kamensky2018Contact,Alaydin2021Updated}. To include self-contact, \emph{finiteness} is achieved by excluding a fixed-size area or sphere in the undeformed state from the integral. However, this exclusion eliminates the guarantee that under all deformations there is no penetration (Figure~\ref{fig:SL-problems}(1)).

To avoid \emph{spurious repulsion}, the potential is localized to a distance less than that of the self-contact exclusion distance. In this way, in the undeformed state, there is no repulsion.  However, if compression is high enough, spurious forces will appear (Figure~\ref{fig:SL-problems}(2)). 

As these potentials are defined as integrals on smooth surfaces and then discretized, there is a high degree of \emph{mesh independence}, especially for finer meshes, but may lead to self-intersection for extreme deformations, as described above.  Furthermore, as pointed out in \cite{Kamensky2018Contact}, the discretization based on pairwise potentials between quadrature points for extreme deformation may also result in failure to satisfy contact constraints (Figure~\ref{fig:SL-problems}(3)). 

\subsection{Repulsive shells \cite{repulsive2024}}  Building on, e.g., \cite{strzelecki2013tangent}, \cite{repulsive2024} introduced a repulsive potential, whose minimization can be used to define self-avoiding surfaces. 
An important feature of this potential in our context is that it handles the problem of near-point interactions gracefully for smooth surfaces, as the potential depends not only on positions but also on the normals. It vanishes in the limit of points approaching a common position on a smooth surface, ensuring \emph{finiteness} for smooth surfaces, but not for surfaces with sharp features, as the normal alignment of close points is needed.  As these potentials are designed to ensure that surfaces minimizing these are $C^1$ (i.e., precisely to eliminate any possible sharp features), they use high inverse powers of distance, resulting in divergence for piecewise smooth surfaces.

 As the interaction potential for any curved surface does not vanish, \emph{spurious repulsion} is present for an undeformed surface (in the context of the target application of \cite{repulsive2024}, defining surfaces as extrema of the repulsive potentials, this repulsion is not spurious).  As the discretization used in \cite{repulsive2024} is based on sampling at quadrature point pairs similar to \cite{sauer2013computational}, it also does not guarantee in general that the contact constraints are satisfied exactly for the discretization, which is particularly important for coarse discretizations.   
 
 We further note that all these potentials depend smoothly on the surface deformation; IPC and \cite{Kamensky2018Contact} are local, and \cite{repulsive2024} is global, although it decays rapidly. 

 \subsection{Barriers based on the gap functions \cite{wriggers2006computational}}
    We briefly mention the possibility of constructing barriers based on gap functions, which can be viewed as solving a standard inequality-constrained contact problem formulation using an interior point method (e.g., \cite{kloosterman2001geometrical}, although this work shifts the barrier towards the interior of an object).

\begin{figure}
    \includegraphics[width=\linewidth]{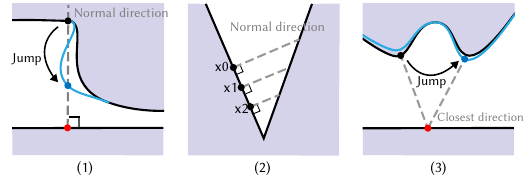}
\caption{(1) The DND gap function is discontinuous when the deformation progresses from the black to the blue curve. (2) The DND gap function can be arbitrarily small for p.w. smooth surfaces as $x_i$ approaches the sharp corner in a continuous setting. (3) The CP gap function changes non-smoothly, as the closest point switches discontinuously (from black to blue). }
\label{fig:gap-function-problems}
\end{figure}

Most contact papers use one of three gap functions:  distance along the normal direction (DND)~\cite{taylor1993finite,vola1998consistent,Hueber2006Mortar,Popp2012Dual,christensen1998formulation,Wang2024Design,kloosterman2001geometrical,Belgacem1998Mortar,Benson1990Single,laursen2002improved}, closest point (CP)~\cite{carpenter1991lagrange,Alart1991Mixed,Simo1992Augmented,Wriggers1995Finite,Pietzrak1991Large,Temizer2014Interior,Taylor1999Smooth,armero1998formulation,Temizer2012ThreeDimensional}, and contact-pair closest distance (CPCD)~\cite{Daviet2020Simple,Li2020IPC,Li2023Convergent,Sassen:2024:RS,HUANG2024GIPC,Shen2024PreconditionedIPC,Otaduy2009Implicit,Verschoor2019Efficient,Macklin2019Nonsmooth,Kim2022Dynamic,Chen2024Vertex,Macklin2020Primal,Razon2023Linear}, the latter defined directly on the discrete surface (e.g., vertex-face and edge-edge distances for PL surfaces). 
Most importantly, the CP gap function does not allow for self-contact without explicit exclusion of nearby points, and, while continuous, is not differentiable with respect to surface deformations (Figure~\ref{fig:gap-function-problems}(3)). \cite{Konyukhov2013} provides a generalized procedure for closest point projection that addresses the issue of uniqueness; however, they don't handle self-contact or complex geometries.
The distance along the normal direction resolves the self-interaction problem, but is discontinuous, hindering the application of efficient numerical methods (Figure~\ref{fig:gap-function-problems}(1)).   While the closest point distance is well-defined for distinct objects if they are not smooth, the distance along the normal direction requires a well-defined surface normal to exclude contact precisely, making it non-trivial to apply for piecewise smooth surfaces.

A potential obtained by applying a logarithmic or inverse power function to these gap functions inherits the properties of these functions. 

Our construction addresses these problems by design, without the need for parameter tuning per simulation: for sufficiently close points, the potential vanishes exactly; no amount of compression leads to potential activation, unless surface points get close to contact; it is defined in a continuum form, so its behavior for different meshes is consistent; its strength is chosen so that it is sufficiently strong for smooth surfaces and does not blow up for piecewise smooth. Finally, we choose the discretization in a way that guarantees that the discretized surfaces are contact-free.

\section{Related Work}
\label{sec:related}

We present a detailed analysis of most of the references discussed here in the supplementary material, summarizing the features of the algorithms and their performance.   We briefly summarize the key aspects here. 

There are many textbooks and reviews on contact mechanics; we focus on the aspects of algorithms most directly related to robustness. \cite{Li2020IPC} supplementary material presents in-depth testing of representative implementations of a number of techniques on a common benchmark, providing empirical evidence of the higher robustness of the barrier-based approach.

\subsection{Barrier/interior-point methods} We group barrier and interior-point methods together, as these are closely related.  These methods, while known for a long 
time (cf. \cite{wriggers2006computational}), only recently became popular in the context of complex deformable contact.  In an early work, \cite{christensen1998formulation}, a direct application of an interior-point method with a non-localized logarithmic potential is compared to a non-smooth Newton method, primarily from the performance perspective. A modified barrier method, with the barrier domain shifted towards the surface interior, and thus not preventing inadmissible configurations, is described in \cite{kloosterman2001geometrical}.

\cite{sauer2013computational} introduced a systematic description of contact methods as surface potentials of several types, including potentials based on repulsion depending on the distance between points, which was extended in \cite{Kamensky2018Contact} and  \cite{Alaydin2021Updated} to handle self-contact (see Section~\ref{sec:related-barrier}). 
\cite{Temizer2014Interior} uses an interior point method to solve the contact problem, but in the presented weak formulation, exact enforcement of the contact is avoided. In \cite{Wang2024Design}, the Ipopt interior point software is used to solve the mortar contact formulation. 

 The barrier method introduced in \citet{Li2020IPC}, to which the discrete version our method is closest,  has been used and extended by numerous follow-up works to include support for codimensional elements~\cite{Li2021Codimensional}, rigid/affine body dynamics~\cite{Ferguson2021Intersection,Lan2022Affine}, medial elastics~\cite{Lan2021Medial}, solid-fluid interactions~\cite{xie2023contact}, higher-order finite element analysis~\cite{Ferguson2023HighOrderIPC}, etc. Most recently, \citet{HUANG2024GIPC,Shen2024PreconditionedIPC} proposed accelerated preconditioned solvers that are amenable to GPU data parallelism to accelerate \cite{Li2020IPC}. These methods sacrifice the exact guarantees of the original IPC to improve performance. 

\citet{Du2023Free},  observes that ``Node-to-Segment'' methods including IPC~\cite{Li2020IPC}) produce spurious tangential contact forces (cf. \cite{Puso2004Mortar}).
To address this, \citet{Du2023Free}  proposes to use a fully $C^1$-continuous surface representation in combination with piecewise-linear meshes,  computing distances used in the contact potential between the discretization nodes and smooth surface, extending the method of \citet{Larionov2021Frictional}. Similar to \cite{Kamensky2018Contact}, \citet{Du2023Free} supports self-collision by ignoring the collision of a vertex with its geodesic neighborhood.  The method does not guarantee contact-free iterations, a feature critical for IPC's robustness~\cite{Li2020IPC}.  Our work is complementary; we do not address this type of artifact in our discretization, a higher-order discretization of our continuum formulation is needed for this. 

Another possible continuum barrier approach to avoiding intersections is to use \emph{tangent-point energies}~\cite{buck1995simple,strzelecki2013tangent} designed to produce self-avoiding smooth surfaces as minimizers. These potentials are global and need to be localized for efficiency and to avoid artificial long-range forces in a physical simulation context. These methods were used in a shape modeling context in   \cite{Yu2021RepulsiveSurfaces,Sassen:2024:RS} with  limitations discussed in Section~\ref{sec:related-barrier}.

In addition to the barrier methods mentioned above,  most notably, \citet{Harmon2009Asynchronous} and \citet{Vouga2011Asynchronous} utilize a set of layered discrete penalty barriers that grow unbounded as the configuration approaches contact. However, this incremental construction makes it unsuitable for optimization-based implicit time integration and therefore requires small time steps for stability. Other notable works include \citet{Kaldor2008Simulating}, which simulates knitted cloth at the yarn level where yarn-yarn collisions are handled through a continuous potential integrating a barrier function over two disjoint spline segments. This approach, similar to surface potentials described above, ignores self-collisions within a single segment and between neighboring segments. Similarly, this work uses quadrature points fixed in the parameter space to compute interactions.

\subsection{Other approaches} The simplest contact response is \emph{penalty-based} \cite{Benson1990Single, Wriggers1995Finite,armero1998formulation,laursen2002improved,Kim2022Dynamic,Chen2024Vertex,Temizer2012ThreeDimensional,MEIER2017972}. Similarly to the barrier methods, the problem is converted to an unconstrained optimization problem with an extra term, but in contrast to barrier methods, a penalty potential is added only if penetration occurs, and no forces are introduced otherwise; by design, penalty methods allow constraint violation, leading to difficult-to-untangle configurations for complex contact. 

\emph{Augmented Lagrangian methods}, widely used in engineering \cite{Alart1991Mixed,Simo1992Augmented,Wriggers1995Finite,Pietzrak1991Large,Puso2004Mortar,Puso2008Segment,hiermeier2018truly,Fernandez2020Topology,Daviet2020Simple,Puso2024Structure,Konyukhov2013}, considerably improve convergence and numerical stability compared to the penalty methods, but also have to go through infeasible configurations before a solution is obtained.  Accurate \emph{mortar methods} \cite{Belgacem1998Mortar,Hueber2006Mortar,Puso2004Mortar} are typically formulated in the augmented Lagrangian form, and do not enforce geometric constraints exactly. 

\emph{"Active set" methods} (which we define broadly as methods that rely in an essential way on identifying a contact set for which the geometric constraints may become equalities, and imposing equality constraints for these v.s. more specific class of active-set optimization methods) include \cite{belytschko1991contact,carpenter1991lagrange,taylor1993finite,vola1998consistent,Hueber2006Mortar,Popp2012Dual}. 
As these methods seek solutions exactly at the boundary of the admissible solution space, having intermediate infeasible solutions is hard to avoid. Many also approximate the exact geometric constraints in the discretization. 

Methods based on reducing the problem to \emph{sequential quadratic programming (SQP)} or \emph{linear complementarity problems (LCP/MLCP)} are widely used \cite{Kaufman2008Staggered,Otaduy2009Implicit,Verschoor2019Efficient,deuflhard2008contact,youett2019globally}. In these approaches the non-linear constrained problem is typically converted to a problem with linear constraints. As a consequence, these algorithms go through inadmissible configurations as a part of the solution process, which affects robustness.  Several works aim to have contact-free states at the end of each time step in a dynamic simulation by resolving interpenetration, but there is no guarantee that this can be achieved~\cite{Kaufman2008Staggered,Otaduy2009Implicit}.

\begin{table*}[htb]

    \caption{
        The table columns correspond to six comparison characteristics: (1) Support self-contact, (2) guarantee intersection-free with conservative CCD, (3) is defined in the continuous setting and discretized after, or directly in the discrete setting, (4) may have spurious rest forces, (5) support types of geometry discretization. We refer to \cref{sec:related} and \cref{sec:eval} for detailed discussions. A more complete comparison with other methods can be found in the supplementary material.}%
    \centering
    \small
    \begin{tabular}{c|c|c|c|c|c|c}
         & Self-contact & Intersection-free & Discrete/Continuous & Localization & Spurious rest forces & Surface types \\ \hline
         \citet{sauer2013computational} & Not descr. & \cellcolor{red!25}No & \cellcolor{green!25}C & \cellcolor{green!25}Yes & \cellcolor{red!25}Yes & PL \\
        \citet{christensen1998formulation} & \cellcolor{red!25}No & \cellcolor{red!25}No & \cellcolor{red!25}D & \cellcolor{green!25}Yes & \cellcolor{green!25}No & PL \\
         \citet{Temizer2014Interior} & \cellcolor{red!25}No & \cellcolor{red!25}No & \cellcolor{green!25}C & \cellcolor{green!25}Yes & \cellcolor{green!25}No & smooth \\
         \citet{Kamensky2018Contact} & \cellcolor{green!25}Yes & \cellcolor{red!25}No & \cellcolor{green!25}C & \cellcolor{green!25}Yes & \cellcolor{red!25}Yes & smooth \\
         \citet{Li2020IPC} & \cellcolor{green!25}Yes & \cellcolor{green!25}Yes & \cellcolor{red!25}D & \cellcolor{green!25}Yes & \cellcolor{red!25}Yes & PL \\
         \citet{Alaydin2021Updated} & \cellcolor{green!25}Yes & \cellcolor{red!25}No & \cellcolor{green!25}C & \cellcolor{green!25}Yes & \cellcolor{red!25}Yes & smooth \\
         \citet{Li2023Convergent} & \cellcolor{green!25}Yes & \cellcolor{green!25}Yes & \cellcolor{green!25}C & \cellcolor{green!25}Yes & \cellcolor{red!25}Yes & PL \\
        \citet{Wang2024Design} & Not descr. & \cellcolor{red!25}No & \cellcolor{green!25}C & \cellcolor{green!25}Yes & \cellcolor{green!25}No & PL \\
        \citet{Sassen:2024:RS} & \cellcolor{green!25}Yes & \cellcolor{red!25}No & \cellcolor{green!25}C & \cellcolor{red!25}No & \cellcolor{red!25}Yes & PL \\
        \citet{HUANG2024GIPC} & \cellcolor{green!25}Yes & \cellcolor{green!25}Yes & \cellcolor{red!25}D & \cellcolor{green!25}Yes & \cellcolor{red!25}Yes & PL \\
        \citet{Shen2024PreconditionedIPC} & \cellcolor{green!25}Yes & \cellcolor{red!25}No & \cellcolor{red!25}D & \cellcolor{green!25}Yes & \cellcolor{red!25}Yes & PL \\
         \citet{Du2023Free} & \cellcolor{green!25}Yes & \cellcolor{red!25}No & \cellcolor{red!25}D & \cellcolor{green!25}Yes & \cellcolor{red!25}Yes & PL + Implicit\\
         Ours & \cellcolor{green!25}Yes & \cellcolor{green!25}Yes & \cellcolor{green!25}C & \cellcolor{green!25}Yes & \cellcolor{green!25}No & p.w. smooth\\
    \end{tabular}
    \label{tab:compare_other}
\end{table*}

\section{Contact Potential Requirements}
\label{sec:requirements}

\begin{figure}[!htb]
    \centering
    \includegraphics[width=\linewidth]{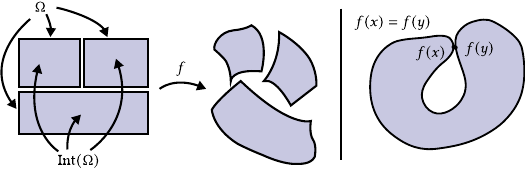}
    \caption{A collection of objects is transformed using a deformation $f$ (left), which can cause the objects to intersect at a contact point (right).}
    \label{fig:domain-notation}
\end{figure}

We consider a collection of deformable objects defined on a material domain $\domain$, whose boundary is $\boundary$. We assume that normals on $\boundary$ are chosen to point outwards.  $\domain$ may have multiple connected components (\cref{fig:domain-notation}).

An admissible deformation 
 $f: \domain \mapsto \mathbb{R}^n$, $n=2,3$ is non-injective on the boundary only, i.e., $f$ is injective in 
 the interior of the domain, but may have \emph{contact points}
$x \neq y$ on the boundary $\boundary$, for which $f(x) = f(y)$.  
We assume there is an initial contact-free map $f_0$ corresponding to the undeformed shape of the object (typically an identity map if $\domain \subset \mathbb{R}^n$).  
 
\begin{figure}[!htb]
    \centering
    \includegraphics[width=0.5\linewidth]{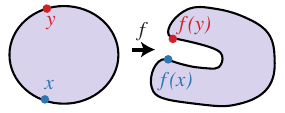}
    \caption{The distance from $x$ to real contact point $y$ may be arbitrarily close, depending on the deformation $f$.}
    \label{fig:arbitrarily-close-points}
\end{figure}

We say that an admissible $f$ is \emph{in contact} if it has contact points, otherwise, we call it \emph{contact-free}. 
We assume that for a point $x \in \boundary$, we have a metric, $d_c(x,f)$, of how far it is from being a contact point. Defining this metric is a key aspect of our construction, and is necessary to formulate our contact requirements. 

While for rigid objects the distance to the closest point on \emph{another} object is adequate to use as $d_c(x,f)$,  as points on the same object cannot move into contact,
 this is not the case for deformable objects. Due to the possibility of self-contact, we cannot exclude points on the same object and, in this case, there are always points on $\boundary$ arbitrarily close to $x$ in Euclidean distance that need to be considered far from being in contact with $x$.  At the same time,
an arbitrarily small, if measured by Euclidean distance, perturbation of a surface may create an actual contact, so the distance from $x$ to a real contact point $y$ along the surface can be arbitrarily small (\cref{fig:arbitrarily-close-points}).

For smooth surfaces, the solution is to take the normals at the points into account. For points in contact, the normals are pointing in opposite directions. So if we include the distance between $n(x)$ and the normals at other points in the definition of $d_c(x,f)$, we can eliminate the nearby points unless the curvature is high, and folds over onto itself, as discussed more precisely in Section~\ref{sec:smooth-surfaces}. For piecewise smooth surfaces, the situation is more complex.

Before describing our approach to the distance to contact, we make the properties we stated in the introduction more precise. 

\begin{requirement}[\Rgen]\label{req:generality}
A total contact potential integral \[\Psi(f) := \iint_{\boundary \times \boundary} \potential(x,y;f)~\mathrm{d}x\mathrm{d}y\] is finite, i.e., the pointwise potential $\potential(x,y;f)$ is integrable on $\boundary \times \boundary$ for piecewise-smooth surfaces $\Omega$ for any $f$ not in contact. $\potential(x,y;f)$ is repulsive, meaning it must be monotonically non-increasing with $d_c(x, f)$ for any $f$.
\end{requirement}

\begin{requirement}[\Rbar]\label{req:barrier} For a time-varying $f_t$,
 $\Psi(f_t)$ increases to infinity for $t \rightarrow t_0$  if the distance to contact $d_c(x,f_t)$ goes to zero for any $x$. Combined with incremental potential time-stepping and \ac{CCD}, this can be used to guarantee that all configurations remain contact-free.
\end{requirement}

\begin{requirement}[\Rrest]\label{req:no-spurious-forces}
Suppose $f_0$ is the initial configuration of the simulation. For any transformation $f$ differing from $f_0$ by a rigid transformation, both $\Psi(f)$ and $\nabla \Psi(f)$ are zero. This requirement is necessary to ensure that the potential does not create artificial forces that would cause motion/deformations if no external forces are acting on the object.
\end{requirement}

\begin{requirement}[\Rloc]\label{req:localization}
$\potential(x,y;f)$ has a locality parameter $\epsilon_\text{trg} > 0$ ($\dhat$ in the notation of \citet{Li2020IPC}), with no restrictions on its magnitude, on which it depends at least continuously, and the potential vanishes if $d_c(x,f) > \epsilon_\text{trg}$. If we solve a sequence of problems with decreasing $\epsilon_\text{trg}$, we approach a solution of the standard inequality-constrained formulation of contact problems.
\end{requirement}

\begin{requirement}[\Rdiff]\label{req:differentiability}
If $f$ is defined by a finite number of parameters (in the simplest case, vertex positions of a mesh) then $\Psi(f)$ depends differentiably, and piecewise twice differentiably, on the parameters of $f$.  Then the potential leads to a force with piecewise continuous Jacobian, allowing for second-order methods for implicit time-stepping. 
\end{requirement}

Importantly, the formulation of most of these properties, except for the first one, can be applied to either the continuum potential or its discretization. The robustness advantages of using a barrier disappear if these are enforced only approximately in the discrete case. For this reason, we add an additional requirement: 

\begin{requirement}[\Rdiscr]\label{req:discretiztion}
    The discretization of the contact potential satisfies requirements \Rgen to \Rdiff (rather than just in the limit of refinement).
\end{requirement}

In particular, this requires constructing a special type of quadrature that ensures the barrier property. 

\paragraph{Remark} The \Rgen requires the integral to be well-defined. Even in the absence of contact, the potential may not be bounded (see Section 4.2 in the supplemental document). In the discrete case, the integral in \Rgen is replaced by a finite sum.

\section{Formulation}\label{sec:contact-candidates}

Our approach to defining contact potentials is conveniently formulated in terms of \emph{{\interaction} sets} $C(x, f)$.  This set is a subset of $\boundary$ away from $x$, for which the barrier potential at $x$ does not vanish.

If we define the pointwise contact potential as 
\begin{equation}
\potential(x,y;f) := \gamma(x,y) \pex(\|f(y)- f(x)\|)
\label{eq:general_psi}
\end{equation}
 where  $\gamma$ is a factor vanishing outside of $C(x, f)$,  $\pex$ vanishes at a distance $\epsilon(x)$. Note that $\gamma$ and $\pex$ have different supports, the intersection of which forms the support of $\potential$. If $\pex$ tends to infinity sufficiently quickly as the argument tends to zero, and $\epsilon(x)$ is chosen so that the potential vanishes for the rest shape, then such {\interaction} sets satisfy \cref{req:barrier,req:localization,req:no-spurious-forces}. In addition, \cref{req:differentiability} may be satisfied by choosing appropriately smooth $\pex$ and $\gamma$.

The definition involves four main components:  the {\interaction} set, the adaptive locality parameter $\epsilon(x)$,  the barrier potential $\pex(x,y)$, and the factor $\gamma(x,y)$, supported on $C(x,f)$. 
As it will be clear from the discussion in the next section, $\gamma(x,y)$ can be viewed as a \emph{direction localization factor}, supplementing the distance localization, to address the problems described in Section~\ref{sec:related-barrier}. 

We start with contact handling for smooth deformable surfaces, then generalize to piecewise smooth surfaces, in each case explaining how these three potential components are defined.

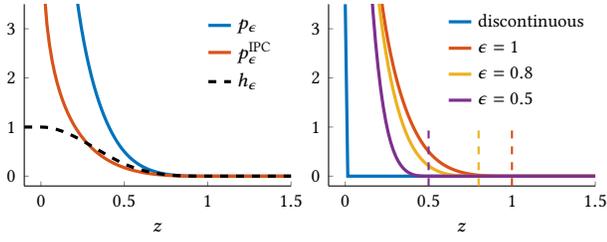
\begin{figure}[tb]
    \centering
    \begin{tikzpicture}[
        declare function={
            B(\x) = (\x <= 0) * 2/3 + and(0 < \x, \x <=1) * (2/3 - \x^2 + 1/2 * \x^3) + and(1 <= \x, \x < 2) * (1/6 * (2-\x)^3);
            he(\x,\epsilon) = B(2 * \x / \epsilon) / B(0);
            b(\d,\dhat) = (\d<0) * 100 +
                 and(0<=\d,\d<=\dhat) * (-(\d-\dhat)^2 * ln(\d/\dhat)) +
                 (\d > \dhat) * 0;
        }
    ]

    \tikzstyle{every node}=[font=\footnotesize]

    \pgfmathsetlengthmacro\MajorTickLength{
      \pgfkeysvalueof{/pgfplots/major tick length} * 0.4
    }

    \pgfplotsset{
        every non boxed x axis/.append style={x axis line style=-},
        every non boxed y axis/.append style={y axis line style=-},
        every axis plot/.append style={very thick},
        legend image code/.code={
            \draw[mark repeat=2,mark phase=2]
            plot coordinates {
                (0cm,0cm)
                (0.15cm,0cm)        %
                (0.3cm,0cm)         %
            };
        }
    }

    \begin{axis}[
        name=axis1,
        width=0.6\linewidth, height=4cm,
        clip mode=individual,
        xmin=-0.1, xmax=1.5, ymin=-0.2, ymax=3.5,
        legend style={
            fill=none,
            draw=none
        },
        legend cell align={left},
        axis lines=left,
        xlabel=$z$,
        no markers,
        ytick={0,1,2,3}, ytick align=inside,
        xtick={0,0.5,1,1.5}, xtick align=inside,
        major tick length=\MajorTickLength
    ]
        \addplot[domain=0:1.5,MATLABColor1,samples=100]{he(x,1)/x};
            \addlegendentry{$\barrier_{\epsilon}$}
        \addplot[domain=-0.01:1.5,MATLABColor2,samples=100]{b(x,1)};
            \addlegendentry{$\pipc$}
        \addplot[domain=-0.1:1.5,black,samples=50,dashed]{he(x,1)};
            \addlegendentry{$\he$}
    \end{axis}

    \begin{axis}[
        at={(axis1.east)}, anchor=west, xshift=5mm,
        width=0.6\linewidth, height=4cm,
        clip mode=individual,
        xmin=-0.1, xmax=1.5, ymin=-0.2, ymax=3.5,
        legend style={
            inner sep=0pt,
            fill=none,
            draw=none
        },
        legend cell align={left},
        axis lines=left,
        xlabel=$z$,
        no markers,
        ytick={0,1,2,3}, ytick align=inside,
        xtick={0,0.5,1,1.5}, xtick align=inside,
        major tick length=\MajorTickLength
    ]
        \addplot[domain=0:1.5,MATLABColor1,samples=100]{(x<=0) * 3.5};
            \addlegendentry{discontinuous}
        \addplot[domain=0:1.5,MATLABColor2,samples=100]{he(x,1)/x};
            \addlegendentry{$\epsilon=1$}
        \addplot[domain=0:1.5,MATLABColor3,samples=100]{he(x,0.8)/x};
            \addlegendentry{$\epsilon=0.8$}
        \addplot[domain=0:1.5,MATLABColor4,samples=100]{he(x,0.5)/x};
            \addlegendentry{$\epsilon=0.5$}
        \addplot +[mark=none,MATLABColor2,dashed,thick] coordinates {(1, -0.2) (1, 1)};
        \addplot +[mark=none,MATLABColor3,dashed,thick] coordinates {(0.8, -0.2) (0.8, 1)};
        \addplot +[mark=none,MATLABColor4,dashed,thick] coordinates {(0.5, -0.2) (0.5, 1)};
    \end{axis}
\end{tikzpicture}
    \caption{\figname{Barriers} Left: plot of the cubic spline $\he(z)$ and barrier function $\barrier_{\epsilon}(z)=\he(z)/z^{n-1}$ for $\epsilon=1$. We also include a plot of the log barrier $\pipc(z)$ of \citet{Li2020IPC} for comparison. Right: our barrier improves approximation to the discontinuous function as $\epsilon$ goes to 0.}
    \label{fig:barrier-p}
\end{figure}

\subsection{Deformable smooth surfaces}
\label{sec:smooth-surfaces}
We first consider the case where both $\boundary$ and $f(\boundary)$ are smooth surfaces, to describe the main ideas without considering the many cases needed for piecewise smooth surfaces.

  Two points of an admissible smooth surface $f$ are in contact, if (a) they coincide in space, $f(x) = f(y)$, and (b) the normals of the point have the opposite orientation, $n(x) = -n(y)$. 
 We define the {\interaction} sets to include points close to contact, i.e., points for which  $f(x)$ and $n(x)$ are both close to  $f(y)$ and $-n(y)$, respectively, and exclude points for which at least one of these requirements is not satisfied. Taking the normals into account addresses the problem of distinguishing between close material points and points close to contact. 

Rather than measuring the distances between normals,   we use a different approach that generalizes naturally to non-smooth points of piecewise smooth surfaces but behaves similarly in the smooth case.  

 \begin{figure}
    \centering
    \includegraphics{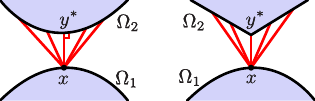}
    \caption{Contact points as a local minimum of distance on the surface.}
    \label{fig:contact-localmin}
\end{figure}

 In \cref{fig:contact-localmin}, observe that for a fixed point $x\in\Omega_1$, if we consider $d_x(y,f) = \|f(y) - f(x)\|$ with $y$ varying over $\Omega_2$, $d_x(y,f)$ has a local minimum at $y^*$, the closest point on the surface. 

If we also consider points $y\in\Omega_1$, i.e., include self-contact, $d_x(y)$ has a local minimum at $x$, but not at any point in a neighborhood of $x$. This leads to the following idea, which is the key to our generalization to piecewise smooth surfaces, as it does not use normals or surface smoothness: 

\begin{quote} \emph{Define the {\interaction} set $C(x,f)$ as points that are close to being local minima of the distance function $d_x(y,f)$ distinct from $x$.}
\end{quote}

Additionally,  we require that the vector $f(y) - f(x)$ points towards the exterior of the surface at $f(x)$ and towards the interior at $f(y)$, which is critical for handling thin objects. 
Next, we derive expressions based on this to define $C(x,f)$ and the factor $\gamma(x,y)$.

\subsubsection{{\Interaction} sets}
For smooth surfaces, the local minima of the distance function from $f(x)$ to $f(y)$ for a contact-free $f$  satisfy the condition 
\begin{equation}
\Phi^m(x,y) := \|(f(y)-f(x))_+ \times n(y)\|   = 0,
\label{eq:phim-smooth}
\end{equation}
where $(\cdot)_+$ denotes normalization to a unit vector. %
The inequality $\Phi^m(x,y) \leq \alpha$, identifies the set of points close to satisfying this condition. We refer to this as \emph{the local minimum constraint}. 

To exclude points for which the vector $(f(y)-f(x))$ points towards object exterior at $y$ define 
\begin{equation}\label{eq:phie-smooth}
\Phi^e(x,y) := -n(y) \cdot ( f(y) -f(x))_+
\end{equation}
Then $f(x)$ is on the side of the outward pointing normal from $f(y)$, if $\Phi^e(x,y) > 0$. We refer to this as 
\emph{the exterior direction constraint}.

\begin{figure}
    \centering
    \includegraphics[width=\linewidth]{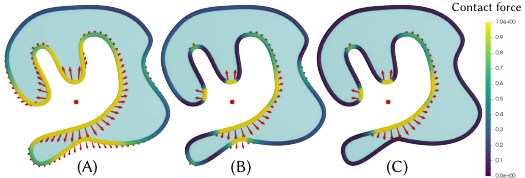}
    \caption{Contact forces (arrows) and potential distribution on the 2D object surface with respect to the red dot. From left to right are: (A) The potential in \citet{Li2020IPC} distributes spherically around the red dot regardless of the surface shape. (B) With local minimum constraints, only surfaces that are close to the local minima or maxima of distance from the red dot have high values. (C) With both local minimum and exterior direction constraints, only surfaces on the closer side of the volumetric object have high values.}
    \label{fig:candidates}
\end{figure}

This leads to the following definition of an {\interaction} set for a smooth surface:
\begin{definition}\label{def:contact-smooth}
The {\interaction} set for a smooth deformation  $f(x)$ with normals $n(x)$ is defined as
\begin{equation}\label{eq:smooth}
C(x,f) := \left\{ y\in\boundary \, \middle\vert \; \begin{array}{l}
    \Phi^m (x,y)  \leq \alpha,  \Phi^e(x,y) \geq -\alpha, \\
    \Phi^m (y,x)  \leq \alpha,  \Phi^e(y,x) \geq -\alpha
  \end{array} \right\}
\end{equation}
where the local minimum constraint $\Phi^m$ and exterior direction constraint $\Phi^e$ are given by \eqref{eq:phim-smooth} and \eqref{eq:phie-smooth}, where $\alpha$ satisfies $0 < \alpha < 1$. We show the impact of $\Phi^m$ and $\Phi^e$ in \cref{fig:candidates}.
\end{definition}
The choice of $\alpha$ is discussed in Section~\ref{sec:eval}; the method works for any choice of $\alpha < 1$, the only impact is on performance, as larger values include more points into the contact set. 
If we define the \emph{distance to contact} $d_c(x,f)$ of a point $x$ under deformation $f$ as the minimum (Euclidean) distance to the {\interaction} set of $x$, which is positive for contact-free surfaces, 
$C(x,f)$ has the following properties: 
\begin{itemize}
\item If $f$ is in contact, $C(x, f)$ contains all contact points of $x$; 
\item If $f$ is not in contact, $d_c(x,f)$ is positive;
\item $d_c(x,f_t)$ tends to zero, as $t\rightarrow t_0$ when a contact-free, time-dependent $f_t$ approaches contact at $t_0$.
\end{itemize}
Please see the supplementary for details.

\paragraph{Remark} In the case of smooth surfaces $(\Phi^m)^2 = 1 - (\Phi^e)^2$, so if $\Phi^e$ is close to 1, $|\Phi^m|$ is close to zero, i.e., the condition of $\Phi^m$ can be ensured by the choice of $\Phi^e$ in \cref{eq:smooth}. However, the relationship is more complex for piecewise smooth surfaces since both normal and tangent are not unique at joints of multiple faces, so we will treat these separately. 

The complete potential for smooth surfaces is given by
\begin{equation}
\Psi(f) =
\iint_{(x,y) \in \boundary^2} \gamma^S(x,y)  p_{\epsilon(x)}(\|f(x)-f(y))\|)~\mathrm{d}x\mathrm{d}y.
\label{eq:potential-deform}
\end{equation}
where $\gamma^S(x,y)$ and $\epsilon(x)$ are defined based on the {\interaction} sets.

\subsubsection{Barrier function} We use
$p_\epsilon(z) := \he(z)z^{-p}$ (\cref{fig:barrier-p}), where the choice of the parameter $p$ is discussed in the supplementary document, we use $p = n-1$, where $n=2,\ 3$ is the dimension of the scene.  Define $\he(z) := \frac{3}{2} B^3\left(2z/\epsilon\right)$, 
a cubic $C^2$ spline basis function with support $|z| \leq \epsilon$ (\cref{app:smoothing}). Then $\he$ vanishes if $d_c(x,f)$ exceeds $\epsilon$, and as a consequence, $\potential(x,y;f)$ is zero. By construction, $\Psi(f)$ becomes infinite if $d_c(x,f) \rightarrow 0$, i.e., this potential satisfies \cref{req:barrier,req:differentiability,req:localization}, as long as the potential grows fast enough. 

\subsubsection{Factor $\gamma^S$} Restricting the potential to the interior of $C(x,f)$ while keeping it differentiable, requires mollification. To construct a suitable $\gamma^S(x,y)$,
 we use a cubic spline basis function for $\Phi^m$, and the smoothed Heaviside step function $H^\alpha(z) := H(z/\alpha)\in C^2(\mathbb{R})$  (\cref{fig:heavyside}, \cref{app:smoothing}) for $\Phi^e$, 

Then we define the {\smoothingfactor} $\gamma^S$ as
\begin{equation}\label{eq:smooth-angle-factor}
\begin{split}
&\delta^\alpha(z) := \frac{2}{\alpha} B^3(\frac{2z}{\alpha}),\\ 
&g^S(x,y) := \delta^\alpha(\Phi^m(x,y)) H^\alpha(\Phi^e(x,y)),\\
&\gamma^S(x,y):=g^S(x,y)g^S(y,x)
\end{split}
\end{equation}
which ensures that the potential is supported within the contact set $C(x,f)$.  Note the similarity between $\he(z)$ and $\delta^\alpha(z)$: one provides localization in distance w.r.t. position of $f(x)$, the other in direction with respect to $n(x)$ for the interaction set points. 

\subsubsection{Adaptive barrier localization.}
To satisfy \cref{req:no-spurious-forces} (\Rrest), we choose  $\epsilon(x)$ for each point $x$ to be equal to 
\[\epsilon(x) := \min( d_c(x,f_0)/2, \epsilon_\text{trg}),\]
where $\epsilon_\text{trg}$ is the global parameter determining the maximal distance at which the potential may be nonzero and $f_0$ is the rest deformation. Note that $\epsilon(x)$ is not necessarily smooth or even continuous as a function of $x$.  This may affect the convergence of the outer integral (under mesh refinement) in computing $\Psi$; however, it does not affect the requirements, in particular, \cref{req:differentiability} (\Rdiff) of the total potential's dependence on the shape parameters of $f$ is still satisfied. %

\begin{wrapfigure}{R}{0.18\textwidth}
    \centering
    \begin{tikzpicture}[
        declare function={
            H(\z) = (\z<=-3) * (0) +
                and(-3 <= \z, \z < -2) * 1/6*(3+\z)^3 +
                and(-2 <= \z, \z < -1) * 1/6*(3 - 9 * \z - 9 * \z^2 - 2 * \z^3) +
                and(-1 <= \z, \z < 0) * (1 + 1/6 * \z^3) +
                (0<\z) * (1);
            Hab(\z,\alpha,\b) = H(3 * (\z-\b) / (\alpha + \b));
        }
    ]

    \pgfplotsset{
        every non boxed x axis/.append style={x axis line style=-},
        every non boxed y axis/.append style={y axis line style=-},
        every axis plot/.append style={very thick},
        legend image code/.code={
            \draw[mark repeat=2,mark phase=2]
            plot coordinates {
                (0cm,0cm)
                (0.15cm,0cm)        %
                (0.3cm,0cm)         %
            };
        }
    }

    \begin{axis}[
        width=4cm, height=3cm,
        clip mode=individual,
        xmin=-0.6, xmax=0.1, ymin=-0.01, ymax=1.1,
        axis lines=middle,
        xtick={-0.5, 0},
        xticklabels={-0.5, 0},
        xlabel=$z$, no markers, ytick={1}
    ]
        \addplot[domain=-1:-0.5,MATLABColor1,samples=2]{0};
        \addplot[domain=-0.5:0,MATLABColor1,samples=20]{Hab(x,1/2,0)};
        \addplot[domain=0:0.5,MATLABColor1,samples=2]{1}
            node[right,pos=1,anchor=north east]{$H^{\alpha}(z)$};
    \end{axis}
\end{tikzpicture}
    \caption{$H^{\alpha}(z)$ with $\alpha=\tfrac{1}{2}$.}
    \label{fig:heavyside}
\end{wrapfigure}

\begin{prop}
The potential  \eqref{eq:potential-deform} satisfies \cref{req:barrier,req:no-spurious-forces,req:localization,req:differentiability,req:generality} if $f$ is a curvature-continuous surface, with $C(x,f)$ given by \cref{def:contact-smooth},  distance-to-contact defined as 
$d_c(x,f) := \min_{y\in C(x,f)} \| f(y) - f(x)\|$, and $\gamma(x,y)$ given by \cref{eq:smooth-angle-factor}.
\label{prop:smooth-properties} 
\end{prop}
The details can be found in the supplementary document.

\subsubsection{Relation to other potentials}  Our potential eliminates many shortcomings of the previously proposed potentials described in Section~\ref{sec:related-barrier}.  

This potential remains finite for admissible smooth surfaces, for our choice of $p =n-1$ (see the supplementary), without excluding a self-interaction radius as in \citet{Kamensky2018Contact}, at the same time retaining the barrier property exactly, thus preventing the possibility of configurations of the type shown in Figure~\ref{fig:SL-problems}.

Spurious repulsion under compression (Figure~\ref{fig:IPC-problems} and \ref{fig:SL-problems}) is eliminated, as these points cannot be in the interaction set, as their normals are parallel or close to parallel.  Similarly, the adaptive choice of $\epsilon$ eliminates spurious forces in the rest state. 

Unlike the IPC potential, the extent of the barrier $\epsilon$ (as well as all other properties) is independent of discretization.

Compared to a barrier constructed from common gap functions as described in Section~\ref{sec:related-barrier}, our potential depends smoothly on the deformation $f$, unlike potentials based on the discontinuous DND or non-smooth CP gap functions. Thus, our potential supports efficient numerical methods. 

Our smooth-surface potential is closest to the tangent energies of \cite{strzelecki2013tangent,Yu2021RepulsiveSurfaces}. Note that the factor $\Phi_m$ can be interpreted as the magnitude of the projection of $f(y) -f(x)$ to the tangent plane of the surface, i.e., exactly the denominator of the energy \cite[~Section 2.1]{Yu2021RepulsiveSurfaces}.  As a consequence, it can be viewed as a form of our potential without the mollification in the factor $\gamma$, i.e., with $\gamma$ set  to $\Phi^m$ and the part related to $\Phi^e$ dropped, and localization in space.  As a consequence, there are repulsive forces for arbitrarily close points on curved surfaces, vanishing only in the limit, since the tangent-point distance is finite for $x$ and $y$ with non-parallel normals. There is also interaction between two sides of, e.g., a very thin object. 

Another important difference is the choice of the power of the potential $p$. In the setting of 
\cite{Yu2021RepulsiveSurfaces}, the power $p$ is set high to ensure that the surfaces determined by the energy are $C^1$, while in our case, these are not determined by the energy minimization, and their smoothness is ensured by other forces (e.g., elasticity). Moreover, finiteness for p.w. smooth surfaces imposes an upper bound on $p$. 

However, we emphasize that in the case of piecewise smooth surfaces (as a more general case of piecewise linear meshes, which are widely used in physics simulations), normals are not uniquely defined, and significant adjustments are needed to the repulsive potentials to apply in this case. 
Moreover, for all desired properties to carry over to the discrete case, the discretization needs to be constructed in a particular way (Section~\ref{sec:discretization}).

\subsection{Piecewise smooth contact}\label{sec:pw-sm}
 We now extend the potential to piecewise smooth surfaces. There are two reasons for this extension: (1) piecewise smooth surfaces are ubiquitous in graphics and scientific computing as they can represent shapes with sharp features, and (2) smooth surfaces are often approximated by piecewise linear surfaces for which efficient and robust continuous collision detection is available.

We consider surfaces consisting of patches $\boundary_i$, which form a possibly curved manifold mesh satisfying the standard definition \cite{do2016differential}. We assume that each patch has continuously varying normals defined everywhere, including boundary (i.e., no cones are allowed), the edge curves meeting at a vertex, and faces meeting at an edge have distinct tangents and tangent planes in the undeformed shape. We refer to the curves separating patches as \emph{edge curves}, or simply edges, and points shared by more than two patches
as vertices.

In this case, there are six possibilities for a contact point, corresponding to the possible pairs of contacts between any two element types: Face-Face, Face-Edge, Face-Vertex, Edge-Edge, Edge-Vertex, Vertex-Vertex. We handle all these cases together in a uniform way.

\subsubsection{{\Interaction} sets for piecewise smooth surfaces}
We use the same general definition for {\interaction} sets $C(x,f)$ for piecewise smooth surfaces. Specifically, \emph{$C(x,f)$ is the set of points $y$ close in the sense defined below to local minima of the distance function $\|f(y)-f(x)\|$, and with the vector $f(y)-f(x)$ pointing to the exterior of the surface at $x$ and to the interior at $y$}. 

However, more complex machinery is needed to convert this to a mathematical definition compared to $\Phi^m$ and $\Phi^e$ functions for smooth surfaces.  Most of the effort required is to express the simple definition above as smooth functions of $f$. 

\paragraph{Local minimum constraint}
Consider the distance $\|f(y) - f(x)\|$ as a function of $y$. If the closest point is in the interior of some patch $\Omega_i$, i.e., the contact point is a face point, then for the gradient w.r.t. $y$, we obtain
\begin{equation}\label{eq:minima}
    \transpose{\nabla_y f(y)}~(f(y)-f(x))_+ = 0
\end{equation}
at the closest point, where the gradient $\nabla_y$ is computed with respect to a parametrization of $\boundary_i$. The columns of the matrix $\nabla_y f$ are two tangents at $y$, i.e., this condition is equivalent to the condition that $n(y)$ is parallel to $f(y)-f(x)$ in the definition of $C(x,f)$.

To generalize the minimum condition to edge and vertex contact points on piecewise smooth surfaces, consider the set of patches $\boundary_i$ containing a point $y$ (one patch for face points, two for edge points, and any number for vertex points). For each $\boundary_i$ and each parametric direction $p\in\mathbb{R}^2$ at $y$, there is a well-defined unit tangent direction $t:=\partial_p f(y)$ of $\boundary_i$. The directional derivative $\partial_p $ is one-sided if $y$ is on the boundary of $\boundary_i$.

The condition for a local minimum is that for any $\boundary_i$, the distance does not decrease along any tangent $t$, i.e.,
\begin{equation}\label{eq:parametric_minima}
    \transpose{t}~(f(y)-f(x))_+  \geq  0
\end{equation}

It turns out to be sufficient to enforce this condition along \emph{three} directions for each patch incident at a point, as shown in Figure~\ref{fig:tangent-notation}, and explained in more detail in Appendix~\ref{sec:pw-smooth-appendix}.

\begin{figure}[!htb]
    \centering
    \includegraphics[width=\linewidth]{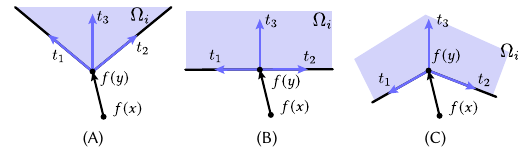}
    \caption{Notation for tangents in \cref{def:contactmin} in 3D. $\boundary_i$ is a face (not necessarily triangle) in 3D colored with light purple, $t_1$ and $t_2$ are the two tangent vectors at $f(y)$, $t_3$ is the angle bisector of the angle between $t_1$ and $t_2$. The face is viewed from the normal direction. 
    }
    \label{fig:tangent-notation}
\end{figure}

\begin{figure}[!htb]
    \centering
    \includegraphics[width=\linewidth]{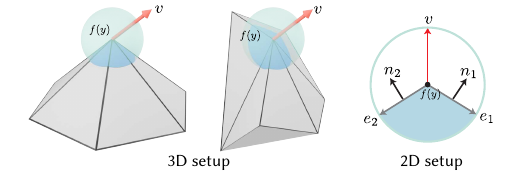}
    \caption{Problem setup of determining if a vector $v$ (red) is pointing inside of the piecewise linear surface. Point $f(y)$ and its 1-ring neighbors on the piecewise linear surface (gray) intersect with a small sphere (green) centered at $f(y)$, their intersection is shown in blue. The intersection in 3D is not necessarily convex (middle).%
    }
    \label{fig:cone-filter-setup}
\end{figure}

\paragraph{Exterior direction constraint}
The analog of $\Phi^e(x,y)>0$ requires a criterion for determining that a unit vector $v:=(f(y) - f(x))_+$ at a point $f(y)$ of a closed piecewise smooth surface points inside. For piecewise linear surfaces, as in \cref{fig:cone-filter-setup}, the point $f(y)$ and its 1-ring neighbors form a cone.  Consider a small sphere centered at $f(y)$, its surface is partitioned into two parts by the cone, and $v$ intersects the sphere surface at one point. The problem is identifying the part of the sphere containing the point. 
For high-order surfaces, determining if $v$ points inside only requires checking the cone formed by the tangent planes at $f(y)$, which is equivalent to considering the same problem for piecewise linear surfaces.

We are going to define a function $\Phi^e(x, y)$ that is positive if and only if $v$ points inwards at $f(y)$.  Consider the problem in 2D (\cref{fig:cone-filter-setup}). The two incident edges at $f(y)$ are $e_1,\ e_2$, with corresponding normals $n_1,\ n_2$. We assume all vectors have unit length since we only care about the directions. Suppose $e_1\times n_1>0$ (otherwise swap $e_1$ and $e_2$), to determine if $v$ points outside the surface we only need to check if $e_1,\ v,\ e_2$ are in counter-clockwise order, so we can define
\begin{equation}\label{eq:cone-filter-2D}
    \Phi^e(x, y):=(v-e_1)\times(v-e_2).
\end{equation}
Note that $\Phi^e(x,y)=0$ if and only if $v,\ e_1,\ e_2$ are on the same line, i.e. $v$ overlaps with $e_1$ or $e_2$.

\begin{figure}[!htb]
    \centering
    \includegraphics[width=0.8\linewidth]{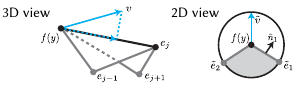}
    \caption{3D (left) and 2D (right) views of the closest point on an edge. The 2D view is the projection of the 3D view onto the plane perpendicular to the edge $e_j$. $\tilde{v}, \tilde{e}_1, \tilde{e}_2$ are the projections of $v, e_{j-1}, e_{j+1}$ respectively. $\tilde{n}_1$ is the projected normal of the face bounded by $e_{j-1},e_j$. Deciding if $v$ points inside the cone in 3D is equivalent to deciding if $\tilde{v}$ is inside the sector bounded by $\tilde{e}_1$ and $\tilde{e}_2$ in 2D.}
    \label{fig:closest-shared}
\end{figure}

In 3D, we first discuss $\Phi^e(x,y)$ for edges in Edge-Vertex and Edge-Edge contact. As shown in \cref{fig:closest-shared}, to determine if a vector $v$ points inwards, we can project everything onto the plane perpendicular to edge $e_j$ and apply \cref{eq:cone-filter-2D} on the projected vectors. For vertices, however, defining $\Phi^e(x,y)$ becomes more complicated. A detailed description of how to define $\Phi^e(x,y)$ in a robust fashion for vertices is included in \cref{app:outside-filtering}. This leads to the following definition of an {\interaction} set:

\begin{figure}
    \centering
    \includegraphics[width=\linewidth]{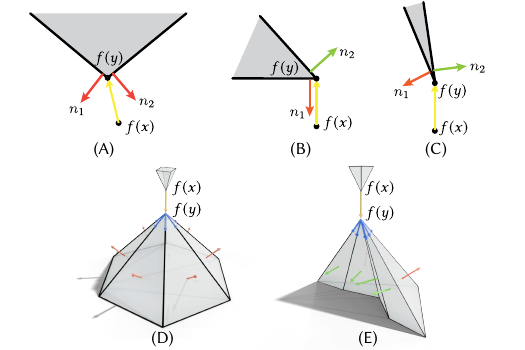}
    \caption{2D (top) and 3D (bottom) normal directions for contact points. $f(y) - f(x)$ is shown in yellow; $n_i(y)$ are shown in red if \cref{eq:normal} is satisfied, otherwise green. Tangent directions are shown in blue in 3D. In simple contact cases, e.g. (A) and (D), the normals all satisfy \cref{eq:normal}, but in other cases, not all normals satisfy the inequality.}
    \label{fig:normal-notation}
\end{figure}

\begin{definition}[{\Interaction} sets for p.w. smooth surfaces]
For each point $y$ of $\boundary$, let $I$ be the set of indices of patches $\boundary_i$ containing $y$, and  for each patch, define $t^k_i(y)$, $k=1,2,3$ as above.  Then $C(x,f)$ consists of points $y$ satisfying
\begin{equation}
\begin{split}
&\Phi^m_{ik}(x,y):= \transpose{t^k_i(y)} (f(y)-f(x))_+  \geq -\alpha,\\
&\Phi^m_{ik}(y, x) \geq -\alpha,\;
\Phi^e(x,y) \geq -\beta,\;
\Phi^e(y, x)  \geq -\beta,\;
\end{split}
\label{eq:pw-contact-alpha}
\end{equation}
for $i \in I$, $k=1,2,3$, $0<\alpha\leq 1$, and $0<\beta\leq 1$ is a separate parameter that controls the smoothness of $\Phi^e$. We use $\beta=0.1$; the specific choice of $\beta$ has little impact on the method's behavior and numerical stability. 
\label{def:contactmin}
\end{definition}

\begin{remark}

The size of the index set $I$ depends on the type of primitive being considered: In 2D, edge curves have size $|I|=1$; vertices have $|I|=2$, including the two edges joined at this vertex. In 3D, faces have $|I|=1$, edge curves have $|I|=2$, and vertices have $|I|\geq 2$.
\end{remark}

Using mollification, we define factors $g^e(x,y)$ and $g^m(x,y)$ from $\Phi^e$ and $\Phi^m_{ik}$ as described in \cref{sec:pw-smooth-appendix}. 

\subsubsection{Contact potential for piecewise smooth surfaces}  Using the new definition of {\interaction} sets $C(x)$ and combining the local minimum and exterior direction constraints, we get, similar to the smooth case, 
\begin{equation}\label{eq:pw-angle-factor}
\begin{split}
&g^{PS}(x,y) := g^e(x,y) g^m(x,y)\\
&\gamma^{PS}(x,y) := g^{PS}(x,y) g^{PS}(y,x)
\end{split}
\end{equation}

However, one cannot simply replace $\gamma^{S}(x,y)$ with $\gamma^{PS}(x,y)$ in \cref{eq:potential-deform} to obtain the potential for the piecewise smooth case.  Recall that contact may happen to measure-zero sets on the piecewise smooth surface, e.g., vertex-vertex and edge-edge contact. Consequently, the contact set in these configurations has zero measure (edge curves and vertices in 3D), and direct surface integration of $\psi(x)$ will lead to a zero potential.

For this reason, we treat low-dimensional elements separately. Informally, we can think about the interaction of lower-dimensional elements, as expanding them into areas: e.g., an area of width $L$ along curves, and an area of size $L^2$ assigned to vertices, with the potential constant along the direction in the area orthogonal to the curve, or on the whole area assigned to a vertex.  More precisely, this corresponds to adding line integrals for edge curves weighted by $L$, and point sums for vertices, weighted by $L^2$.  The potentials we integrate for a pair of elements $G$ and $H$, possibly of different dimensions, e.g., with $G$ being a vertex and $H$ a face,  can be written in the form
\[
P(x,y; G, H) = \gamma^{PS}(x,y; G, H)\pe(x,y)
\]
where $\gamma^{PS}(x,y; G, H)$ is the factor defined in \cref{eq:pw-angle-factor}, \emph{for $x$ considered as a point on element $G$,} in other words, if $G$ is a face, and $x$ happens to be an edge or vertex point, the factor $\gamma^{PS}(x,y)$ is computed using formulas for face points $x$.

Then the total potential can be written as
\begin{equation}\label{eq:pwpot}
\Psi(f) = \sum_{ (g,h) } \sum_{\substack{i \in I_g, j \in I_h\\ G_i \cap H_j = \varnothing}} L^{4-\dim g - \dim h} \int_{G_i}\int_{H_j} P(x,y; G_i, H_j)~\mathrm{d}x\mathrm{d}y
\end{equation}
where $g,h$ are one of the element types $\{\text{Face}, \text{Edge}, \text{Vertex}\}$, with the first sum is over 9 unordered pairs, $I_g$ and $I_h$ are sets of indices of elements of types $g$ and $h$, and $G$ and $H$ element indices.  The integrals are area, line, or 0-dimensional, i.e. simple evaluations for vertices. %

\begin{prop}
The potential  \eqref{eq:pwpot} satisfies \cref{req:barrier,req:no-spurious-forces,req:localization,req:differentiability,req:generality} if $f$ is piecewise-curvature-continuous surface as defined in this section, with $C(x,f)$ given by \cref{def:contactmin},  distance-to-contact defined as 
$d_c(x,f) := \min_{y\in C(x,f)} \| f(y) - f(x)\|$, and $\gamma(x,y)$ given by \cref{eq:pw-angle-factor}.
\label{prop:pwsmooth-properties} 
\end{prop}
Please see the supplementary document for details.

\subsubsection{Parameters of the potential} Whether the potential satisfies contact potential requirements does not depend on the choice of $L > 0$, $\epsilon_\text{trg} > 0$,
and $0 < \alpha < 1$, as long as these are positive. There is no strong impact on robustness, as long as the values are not too close to the bounds,
but it may affect performance. 

\begin{itemize}
    \item $\alpha$ determines the size of the {\interaction} sets, and as a consequence, how smoothly the potential depends on $f$. The closer $\alpha$ is to zero, the less smooth the dependence is, but the more localized it is.
    \item $\epsilon_{trg}$ determines the upper bound on how far the potential extends ($\hat{d}$ in IPC), but especially at concave corners, $\epsilon_{trg}$ as well as $f_0$, affect this.
    \item $L$ %
    determines the strength of potential for low-dimensional contact. In our examples, for vertices, we set it to the average edge length around the vertex; for edges, we set it to the length of the edge; for faces, $L$ is not needed.
   
\end{itemize}

\subsection{Discretization}
\label{sec:discretization}
For a barrier method to guarantee that the geometric constraints
are not violated, and maintains the attractive features of the potential, we must ensure that the discrete version satisfies the same conditions, most importantly:
\begin{itemize} 
\item requirement~\ref{req:barrier} (the exact barrier property with respect to the discretized geometry); 
\item requirement~\ref{req:differentiability} (differentiability). 
\end{itemize}

It turns out that these requirements are, to an extent, conflicting, and additional effort is needed to satisfy both.  We consider the lowest order discretization for piecewise-linear surfaces, leaving extensions to higher orders as future work.  Designing a discretization that, on the one hand, converges to the underlying continuum potential and, on the other hand, satisfies all requirements is nontrivial already in this case. 

\subsubsection{Ensuring the barrier property} If we use standard quadrature points as it is done in \cite{Kamensky2018Contact,Yu2021RepulsiveSurfaces}, 
the integral approximations in Equation \eqref{eq:pwpot} may have optimal accuracy, but the barrier property is satisfied approximately, and interpenetration is possible, as discussed in \cite{Kamensky2018Contact} and Section~\ref{sec:related}. 

To address this problem, \emph{we choose the closest points on element pairs as our quadrature points}, possibly reducing integration precision, but still satisfying the requirements. 

In this case, the lowest-order discretization of the potential integral is 
\begin{equation}
\Psi^D(f) = \sum_{ (g,h)} \sum_{(i,j)} L^{4-\dim g - \dim h} P(x_i,y_j; G_i, H_j) A(G_i) A(H_j)   
\label{eq:discretization}
\end{equation}
where $(x_i,y_j)$ is a pair of closest points on elements $G_i$ and $H_j$, respectively, and $(G_i,H_j)$ are pairs of non-adjacent elements, and $A(G_i)$ and $A(H_j)$ are measures of elements (area for faces, length for edges and 1 for vertices). 

Different from \cref{eq:pwpot}, the sum above over $(g, h)$ only includes four types: Vertex-Face, Edge-Edge, Vertex-Edge, Vertex-Vertex, since in the case of piecewise linear surfaces, other types are reduced to these types when only the closest points are considered. \cite{Li2020IPC} only considers Vertex-Face and Edge-Edge, since it is sufficient to satisfy \cref{req:barrier} in their case. In our work, due to the mollification discussed below, we include Vertex-Edge and Vertex-Vertex. In practice, we do not have more contact pairs than \cite{Li2020IPC} (\cref{tab:n_contact}), since our contact set is more narrow due to our local minimum and exterior direction constraints.

As the closest distances between points are used, the terms in Equation \eqref{eq:discretization} approximating corresponding terms in Equation 
\eqref{eq:pwpot}, bound them strictly from above, ensuring the exact barrier property.  One exception is the terms for adjacent elements, for which the closest distance is zero, and the bound would be infinity. 

From the perspective of convergence of the potential to the limit under refinement, this is not an issue, as the fraction of adjacent pairs among all pairs converges to zero under refinement.  Excluding interactions between these elements does not affect the barrier requirement either: for piecewise linear meshes, if adjacent elements are in contact, other, nonadjacent elements will be in contact (see Appendix~\ref{sec:adjacent-exclusion} for a more detailed discussion), i.e., \cref{req:barrier} still holds.

However, in the piecewise linear case, the contact between these elements is possible only if a non-adjacent pair is in contact, so these can be safely omitted.

However, using the closest points as quadrature points creates a new complication: 3D positions of quadrature points with a fixed position in the material space naturally depend smoothly on the deformation $f$, but this is not the case for the positions of closest points. 

\subsubsection{Ensuring differentiability}
For the discrete potential to depend differentiably on the shape parameters (vertex positions in the piecewise linear case), we need an additional modification.  First, observe that the distance between two elements is already $C^1$ with the only exception being the distance between two parallel edges, which can be mollified as explained in \cite{Li2020IPC}, so the term $\pe$ in $P$ does not require modifications.  However, the \smoothingfactor $\ \gamma^{PS}$ \cref{eq:pw-angle-factor} depends on the direction between closest points, which may be only $C^0$ continuous with respect to the deformation $f$, when the closest point is on the boundary of the face or edge (\cref{fig:nonsmooth-closest-point}). We introduce a mollification $M(x,y)$ for the closest point pair direction, as an extra multiplicative factor for $\gamma^{PS}$:
\begin{equation}\label{eq:pw-angle-factor-with-mollifier}
\gamma^{PS}(x,y) := g^{PS}(x,y) g^{PS}(y,x) M(x,y),
\end{equation}
 Importantly, introducing this factor does not affect other requirements for the potential.  The details of the construction of this factor 
 can be found in Appendix~\ref{sec:distance-mollification}.

\subsubsection{\Smoothingfactor: local minimum terms}
\begin{figure}
    \centering
    \includegraphics[width=\linewidth]{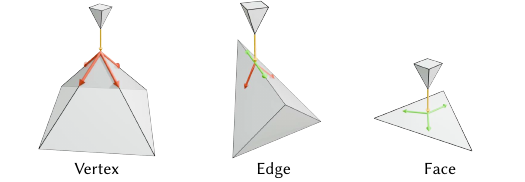}
    \caption{Local minimum constraints for closest points in contact. $f(y)-f(x)$ is shown in yellow, tangent directions included in  $\Phi^m$ are in red, otherwise green. By the property of the closest point, the distance function $d_x(y)$ has zero derivatives along the green tangents, so \cref{eq:pw-contact-alpha} is always satisfied.}
    \label{fig:tangent-demo}
\end{figure}

In the case of only the closest points being used as quadrature points, we can simplify the local minimum terms discussed in \cref{sec:pw-sm}. If a face point $y\in\boundary_i$ is the closest point on $\boundary_i$ to a fixed point $x$, then $y$ is a local minimum of the distance function $d_x(y)=\|f(y) - f(x)\|$, i.e. \cref{eq:minima} is satisfied, so no local minimum constraint is needed. Similarly, if $y$ is on an edge, then $y$ is a local minimum of $d_x(y)$ on the edge, so the local minimum term along the edge can be ignored (\cref{fig:tangent-demo}).

\subsubsection{Adaptive local factor $\epsilon$}
We introduce adaptive $\epsilon$ for our formulation. We first specify a fixed $\epsilon_{trg}$ for the simulation and collect contact pairs in the rest configuration within $\epsilon_{trg}$ and with nonzero potential values. We pick $\epsilon$ for every primitive (vertex/edge/face) so that none of the contact pairs are active in the rest configuration. Then the $\epsilon$ for every edge (face) is chosen to be the minimum among its neighboring vertices (edges). Benefiting from local minimum and exterior direction constraints, the adaptive $\epsilon$ in our method can be much larger than in \cite{Li2020IPC} (\cref{fig:dhat-distribution}).

\subsubsection{Satisfying requirements} Just as the smooth and piecewise smooth potentials, the discretized potential \ref{eq:discretization} satisfies our requirements:

\begin{prop}
   The discrete potential \ref{eq:discretization}, with the {\interaction} sets and {\smoothingfactors} computed as described above, satisfies the 
    \cref{req:barrier,req:no-spurious-forces,req:localization,req:differentiability}. 
\end{prop}
We note that there is no need for \cref{req:generality} for a discrete potential, as it is a finite sum: \emph{any} potential that grows to infinity for $\|f(y) - f(x)\| \rightarrow 0$ can be used, as this is a finite sum, and, unlike integrals, it is unbounded when any term is unbounded.  As we use the closest points on elements as quadrature points, the barrier property is guaranteed in the discrete case. 
 
However, if the growth rate of the potential is too low for the smooth or piecewise-smooth integral potentials, this means that under refinement, the potential will become progressively weaker (\cref{fig:corner-hit}).

The remaining properties are verified in a similar way to the continuum case (see the supplementary 
document). 

\subsubsection{Convergence under refinement} Above, the piecewise smooth formulation is used directly to discretize the potentials on piecewise linear meshes.  A natural question is whether the potential for piecewise linear meshes sampled from a smooth (or piecewise smooth) surface will converge to the potential directly defined for this surface.  Mathematical analysis of convergence is beyond the scope of this paper, but we expect that the piecewise linear mesh potentials converge to the potential defined on the smooth mesh if the parameter $L$ is adjusted in the same way as the mesh edge lengths, i.e., decreased by a factor of two if the mesh is refined uniformly.  Intuitively, the discretization corresponds to sampling the integrand at a set of points on faces, edges, and vertices and weighting by areas associated with them ($L$ controls the size of areas assigned to edges and vertices), which suggests this scaling of $L$.  

The situation is more complicated for the refinement of piecewise smooth surfaces. In this case, the limit surface integral contains the low-dimensional terms (Edge-Edge, Vertex-Face, etc.). These are computed on sharp feature curves and at vertices embedded in the surface, and in discretization corresponding to piecewise linear feature curves and vertices embedded in a piecewise linear mesh.  In this case, for convergence to the correct limit potential, the factors $L$ for edges and vertices on these feature curves should \emph{not} be adjusted, unlike factors for edges and vertices inserted on faces.

\subsection{Friction}
\label{sec:friction}
The friction formulation in \cite{Li2020IPC} can be reused in our setting with minor modifications, which we report here for completeness.
In IPC \cite{Li2020IPC}, the friction force for a contact pair is
$$
F_k(x)=-\mu\lambda_k T_k f_1(\|u_k\|) \frac{u_k}{\|u_k\|},
$$
where $\mu$ is the friction coefficient, $T_k(x)$ is the local tangential basis, $\lambda_k$ is the contact force magnitude, $u_k$ is the local relative sliding displacement at contact $k$,
$$
f_1(y)=\begin{cases}
    -\frac{y^2}{\epsilon_v^2 h^2} + \frac{2y}{\epsilon_v h},& y\in(0,h\epsilon_v)\\
    1,& y\geq h\epsilon_v,\\
\end{cases}
$$
where $h$ is the time step size, $\epsilon_v$ is a velocity magnitude bound that controls how accurately the friction is approximated. Then the dependency of $F_k(x)$ on $T_k$ and $\lambda_k$ is made explicit (or lagged), to obtain the friction potential
$$
D_k(x)=\mu \lambda_k^n f_0(\|u_k\|).
$$
where $\lambda_k^n$ is the contact force magnitude from a prior nonlinear solve (or previous time step) $n$. Since the dependency of $D_k(x)$ on $\lambda_k^n$ is explicit, $\lambda_k^n$ can be directly replaced by the contact force magnitude in our method.

\begin{figure}[!htb]
    \centering
    \includegraphics[clip,width=\linewidth]{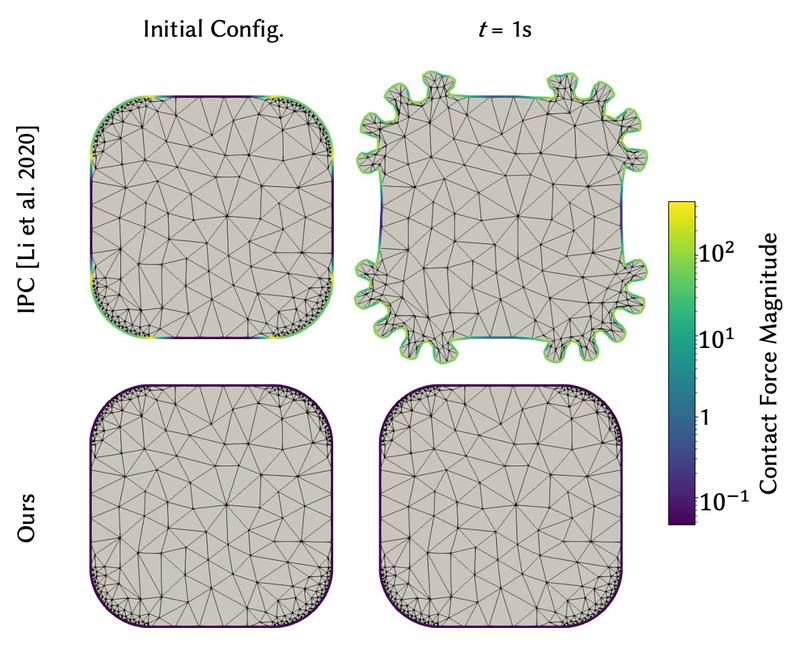}
    \caption{\figname{Spurious stresses: nonuniform squircle mesh} %
    Our method properly handles nonuniform meshes. In this case, we consider a 2D rounded block (of size $\qty{1}{\m}$ x $\qty{1}{\m}$) with maximal edge length $\qty{0.21}{\m}$ in its straight sections and minimal edge length $\qty{0.01}{\m}$ at its corners. Top row: with $\hat{d}=\qty{0.1}{\m}$, IPC introduces spurious contact forces in the refined corners, resulting in a deformation. Bottom row: our method avoids this by filtering based on tangent directions, without the need for small $\epsilon(x)$.} %
    \label{fig:spurious-stresses-nonuniform-2D}
\end{figure}

\section{Evaluation}\label{sec:eval}

We implement our algorithm by extending the IPC Toolkit~\cite{IPCToolkit} (details provided in the supplemental document) \DP{make sure to upload!}. We use Eigen~\cite{Eigen} for linear-algebra operations, Pardiso~\cite{Alappat2020Recursive,Bollhofer2020State,Bollhofer2019Large} for solving linear systems, and PolyFEM~\cite{PolyFEM} as the finite element simulation framework. All experiments are run on a system with an AMD Ryzen Threadripper PRO 3995WX 64-Cores (limited to 16 threads) and \qty{440}{\gibi\byte} of memory. For comparisons against IPC~\cite{Li2020IPC} and ``Convergent IPC''~\cite{Li2023Convergent}, we use the open-source implementations provided in the IPC Toolkit (together with PolyFEM). Please see our supplemental video for animations. All simulation parameters and statistics are summarized in \cref{tab:params}. Our reference implementation, used to generate all results, will be released as an open-source project.

\subsection{Resolving Spurious Forces}\label{sec:spurious-eval}

In this section, we consider a variety of unit tests where the original IPC formulation introduces spurious forces.

\subsubsection{Using larger $\epsilon_\text{trg}$ } 
\begin{figure}[!htb]
    \centering
    \includegraphics[clip,width=\linewidth]{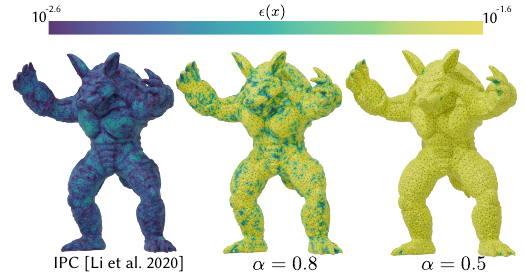}
    \caption{\figname{$\epsilon$ distribution} We implement an adaptive locality parameter with IPC and compare $\epsilon$ after adaptation to avoid spurious rest forces with $\epsilon_{\text{trg}} = \qty{0.02}{\m}$. IPC (left) requires small $\dhat$ all over the surface due to the requirement that $\dhat$ be smaller than the shortest adjacent edge. Due to local minimum and exterior direction constraints, our method (middle and right) is able to concentrate $\epsilon$ refinement on the fingers and toes of the figure.}
    \label{fig:dhat-distribution}
\end{figure}

\begin{figure}[!htb]
    \centering
    \includegraphics[clip,width=\linewidth]{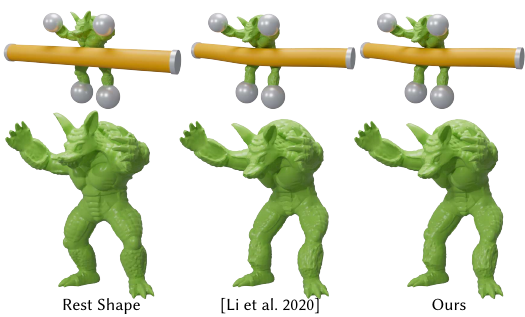}
    \caption{\figname{Armadillo-Bar} An armadillo interacts with an elastic bar. The two ends of the bar are forced to move towards the back of the armadillo, and both the hands and feet of the armadillo are fixed. Both the armadillo and bar are elastic. We show the full scene on the top row, and only the armadillos on the bottom row.}
    \label{fig:armadillo-sim}
\end{figure}

Suppose the input mesh is relatively fine, either globally (a dense mesh) or locally (adaptively refined e.g. to resolve fine features). Having a large $\epsilon_\text{trg}$ allows one to solve the implicit time-stepping problem faster because the barrier is less numerically stiff, and \ac{CCD} needs to do less work in the line search if the objects are kept further apart by the potential. 
We provide an example of this scenario in \cref{fig:spurious-stresses-nonuniform-2D}, where it is natural to have small edges around the rounded corners of a square and long edges along the sides. However, doing so restricts the range of usable $\dhat$ (IPC's notation for $\epsilon_{\text{trg}}$) for the IPC barrier. Using a value ($\dhat=\qty{0.1}{\m}$ in this case) larger than the minimum edge length ($h_{\min}=\qty{0.01}{\m}$) results in spurious forces along the corners and artifacts upon simulation. In contrast, our utilization of an adaptive $\epsilon$ allows us to choose a starting $\epsilon_\text{trg}$ that results in zero initial contact force. %
These rest forces can be avoided while still using a large $\epsilon$ for the non-refined regions (\cref{fig:dhat-distribution}). Further, in \cref{fig:armadillo-sim}, we run a simulation on the same armadillo model with IPC and our method. We choose $\dhat=\qty{0.001}{\m}$ for our method and $\epsilon_{\text{trg}}=\qty{0.0004}{\m}$ for IPC so that there is no contact force at rest shape. Both methods produce similar results, our method takes 512 iterations and 55 minutes (2/3 less than IPC), while IPC takes 1555 iterations and 161 minutes.

\subsubsection{Large Deformation} 
\begin{figure}[!htb]
    \centering
    \includegraphics[width=\linewidth]{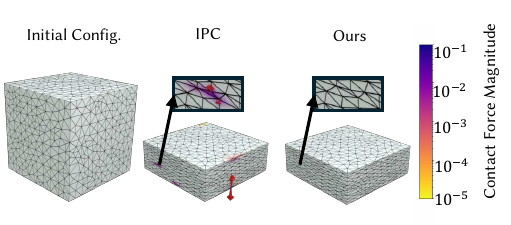}
    \includegraphics[width=\linewidth]{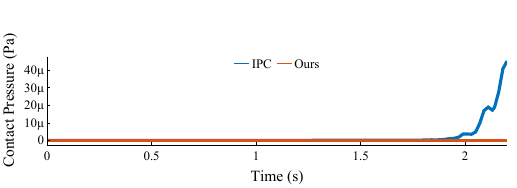}
    \caption{\figname{Spurious stresses: compression} An initially valid choice for $\hat{d}$ can lead to spurious contact (arrows and color) when using IPC. Left: cube at rest with a minimum edge length of $\qty{0.03}{\m}$. Center: upon compression with $\hat{d} = \qty{0.0125}{\m}$, IPC introduces spurious contacts. Right: our method avoids this by considering the tangent directions when finding \interaction sets. Bottom: IPC introduces artificial contact pressure.}
    \label{fig:spurious-stresses-compressed}
\end{figure}

In the presence of large deformations, elements may shrink a lot, and in this scenario, the original IPC barrier formulation adds spurious forces which make the material locally stiffer (\cref{fig:spurious-stresses-compressed}). Our formulation does not suffer from this issue.

\begin{figure}[!htb]
    \centering
    \includegraphics[clip,width=\linewidth]{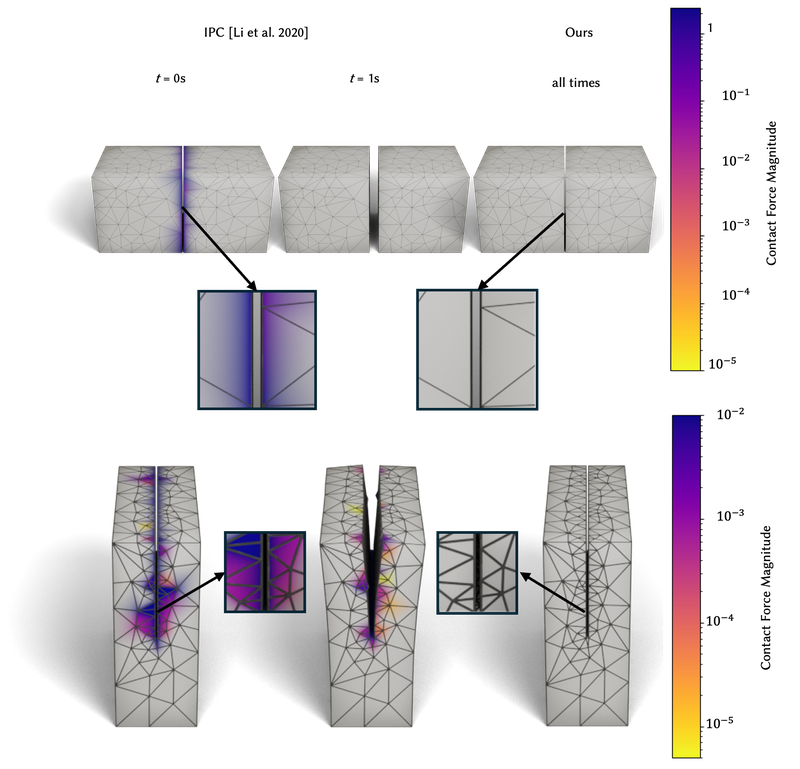}
    \caption{\figname{Spurious stresses: rest configuration 3D} Top row: two cubes initially separated by less than $\hat{d}=\qty{0.025}{\m}$. IPC artificially repels the two blocks while ours does not. Bottom row: a block with a slit of width less than $\epsilon_\text{trg}=\qty{0.0125}{\m}$. Our method does not introduce spurious forces across the slit.}
    \label{fig:spurious-stresses-rest-3D}
\end{figure}

\subsubsection{Spurious stress in the rest configuration} Even if $\dhat < h_{\min}$ is satisfied, the IPC formulation may still have spurious stress in the rest configuration. We show two such scenarios in \cref{fig:spurious-stresses-rest-3D}.

In the first case (\cref{fig:spurious-stresses-rest-3D} Top), two blocks in the initial configuration sit on the plane with the initial distance between blocks less than
$\dhat$. With IPC, they incorrectly start sliding apart without external force applied where the blocks should stay still at all times.
In the second case (\cref{fig:spurious-stresses-rest-3D} Bottom), IPC causes the slit to expand at the top without external forces and stress appears at the bottom.

Our approach avoids spurious forces between close objects by using an adaptive $\epsilon_\text{trg}$. For the contact between primitives on the same object, due to our local minimum and exterior direction constraints, we can use $\epsilon_\text{trg}$ larger than the edge lengths without activating contact at rest configuration (with a wide range of $\alpha$).

\begin{figure}[!htb]
    \centering
    \includegraphics[width=\linewidth]{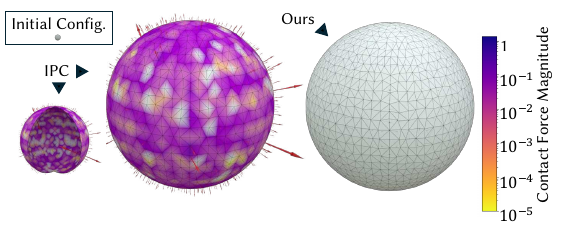}
    \includegraphics[width=\linewidth]{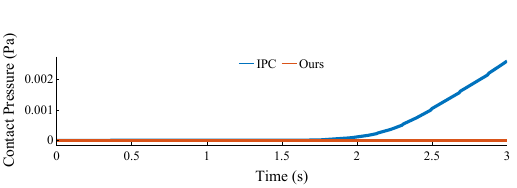}
    \caption{\figname{Spurious stresses: expansion} Upper left: an initially deflated balloon with rest thickness of $\qty{0.1}{\m}$ is expanded by an outward force of $\qty{5}{\kilo\N}$. Center: with $\hat{d} = \qty{0.02}{\m}$, the balloon eventually becomes thin enough that spurious contact forces (represented as red arrows) between its inner and outer layers appear. Lower left: a cross-section showing spurious forces on both the inner and outer layer of the balloon. Right: our method can inflate the balloon until it is arbitrarily thin without introducing artificial contact forces. Bottom: artificial contact pressure is introduced by IPC.}
    \label{fig:balloon}
\end{figure}

\begin{figure}[!htb]
    \centering
    \includegraphics[width=\linewidth]{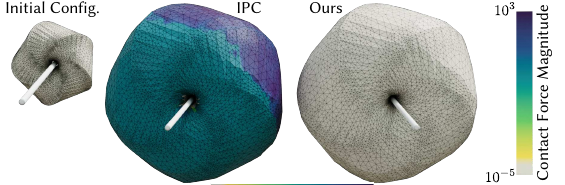}
    \caption{\figname{Spurious stresses: donut expansion} Left: a donut with rest thickness of $\qty{0.02}{\m}$ is inflated around a thin stick by a pressure boundary condition of $0.1t\ \unit{\mega\pascal}$, with simulation time $t$ up to $\qty{0.6}{\second}$. Middle: with $\hat{d} = \qty{0.01}{\m}$, the balloon eventually becomes thin enough that IPC introduces spurious contact forces between its inner and outer layers. Right: our method can inflate the balloon until it is arbitrarily thin without introducing artificial contact forces while still capturing contact forces between the rod and balloon.}
    \label{fig:balloon-donut}
\end{figure}

\subsubsection{Spurious stress under deformation}
Even for a well-chosen $\dhat$ parameter and a benign rest configuration, the IPC barrier can still add spurious forces upon deformation.%

In \cref{fig:spurious-stresses-compressed}, we compress a cube mesh to 33\% of its original height. We assign a Poisson ratio of 0 to the cube to show an example where no bulging and/or folding of the surface occurs. The value for $\dhat$ is initially chosen such that no points are in contact, but upon compression distances shrink and IPC introduces spurious contact forces on the sides of the cube (despite them being flat). Our formulation does not have spurious forces because we use the angle between the direction to the point and surface normal to build our {\interaction} sets.

In \cref{fig:balloon}, we inflate a spherical balloon modeled as a volumetric membrane of thickness $\qty{0.1}{\m}$, hanging on a rigid stick.\footnote{While one could model the balloon using co-dimensional shell elements to avoid this issue, modeling the thickness may be important for analysis or design. For example, with constant outward pressure, the balloon oscillates in thickness, which a shell model would not capture.} As the balloon inflates, its walls get thinner and eventually they become thinner than the initially chosen $\dhat=\qty{0.02}{\m}.$ At this point, IPC treats the inner and outer sides as in contact, introducing forces between the two sides. Our {\interaction} set for one side does not include the other as it accounts for the angle of contact, avoiding this issue. As a more complex example (\cref{fig:balloon-donut}), we inflate a donut-shape balloon around a thin stick. Our method only has contact forces around the center, while IPC has spurious contact forces everywhere. Although one can filter the {\interaction} set by the surface connected component for IPC in these two simple examples (since the inner surface of the shell does not have contact with the outer surface), in more complex scenes (\cref{fig:mat-twist}) it does not work anymore, while our method avoids these issues by relying on our definition of {\interaction} sets.

\begin{figure}[!htb]
\centering
\includegraphics[width=\linewidth]{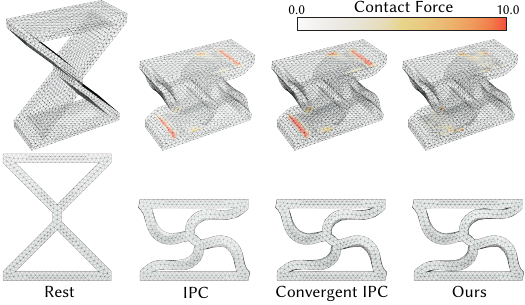}
\caption{\figname{Cusp compression} Compression of a $\chi$-shaped structure. We visualize the contact force distribution on the top row and the front view on the bottom row. Our method significantly reduces the contact forces around the cusp compared with IPC and convergent IPC.}
\label{fig:unit-cell}
\end{figure}

In \cref{fig:unit-cell}, we extrude the extended $\chi$-shaped structure from \cite{joodaky2020mechanics} to 3D and simulate its compression.  When the shape buckles it forms \emph{cusp} contacts at the corners. Since the barrier potentials in different methods have very different scales, for a fair comparison, we pick $\kappa$ for each method so that the minimum distance in the simulation is roughly the same ($\kappa=10^4$ for Convergent IPC and our method, $\kappa=10^5$ for IPC). We show that both Convergent IPC~\cite{Li2023Convergent} and IPC~\cite{Li2020IPC} exhibit large contact forces in these regions, while our method has much less contact forces than both methods. Our method reduces the contact forces around the \emph{cusp} because of the local minimum and exterior direction constraints -- only when edges are close enough to parallel, i.e. the angle of the \emph{cusp} becomes small enough, the contact is activated. Without this constraint, spurious stresses appear in both IPC and Convergent IPC.

\subsection{3D Examples}

\begin{figure}[!htb]
    \centering
    \includegraphics[clip,width=\linewidth]{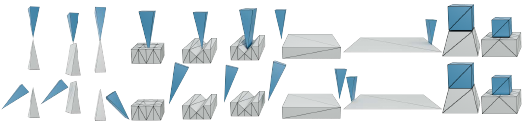}
    \caption{\figname{Erleben tests} We reproduce the test-cases of~\citet{Erleben2018Methodology}. Top: initial conditions involving challenging exact point-point, point-edge, and edge-edge collisions. Bottom: as in \cite{Li2020IPC}, our approach robustly passes all the tests.}
    \label{fig:erleben}  
\end{figure}

\begin{figure}[!htb]
    \centering
    \includegraphics[width=\linewidth]{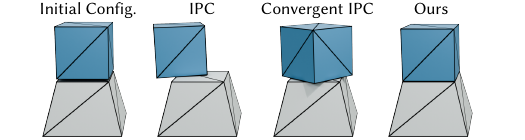}
    \caption{\figname{Erleben test: cliff edge} We reproduce the cliff edge test-case of~\citet{Erleben2018Methodology} with $\epsilon_\text{trg}=\qty{0.01}{\m}$. IPC\cite{Li2020IPC} and Convergent IPC\cite{Li2023Convergent} pass the test but introduce spurious horizontal forces, causing the top block to rotate or slide. Our method significantly reduces extra sliding due to the restriction in contact forces.}
    \label{fig:erleben-cliff-edges}
\end{figure}

We reproduce challenging 3D simulation examples of~\cite{Li2020IPC}. First, we validate our method on the 3D unit tests proposed by \cite{Erleben2018Methodology} in \cref{fig:erleben}. We see similar results to those shown in~\cite{Li2020IPC}, but we highlight one improved result in~\cref{fig:erleben-cliff-edges} where we see reduced spurious tangential movement compared to \cite{Li2020IPC} and \cite{Li2023Convergent}.

\begin{figure}[!htb]
    \centering
    \includegraphics[clip,width=\linewidth]{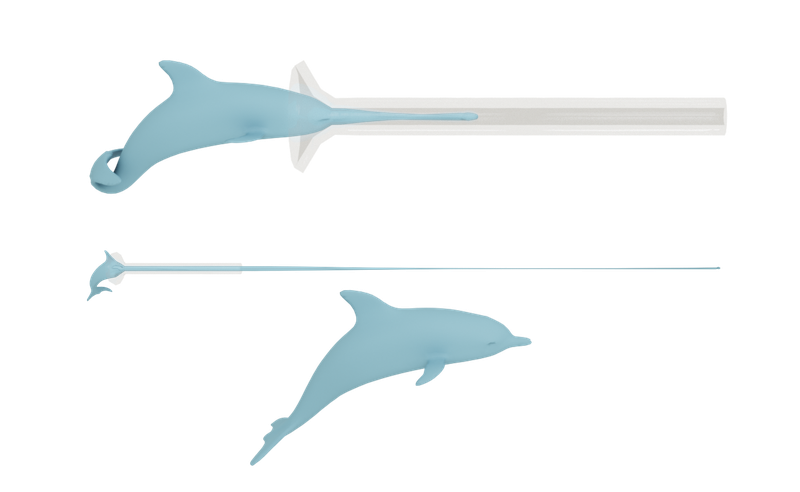}
    \caption{\figname{Dolphin in a funnel} We reproduce the funnel test from~\cite{Li2020IPC} using our method. Top: an elastic dolphin is pulled through a small tube. Middle: this causes extreme deformations. Bottom: the dolphin squeezes through without artifact and recovers its original shape.}
    \label{fig:dolphin}
\end{figure}

\begin{figure}[!htb]
    \centering
    \includegraphics[width=\linewidth]{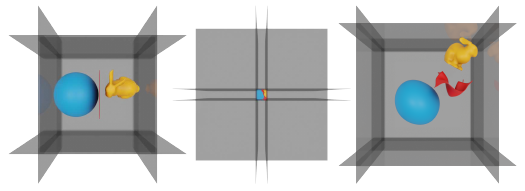}
    \caption{\figname{Trash-Compactor} We reproduce the trash-compactor example from~\cite{Li2020IPC} using our method. Left: three objects are placed in a compactor. Middle: the objects are compressed. Right: the compactor releases and the shapes return to their original shape without intersections.}
    \label{fig:trash-compactor}
\end{figure}

\begin{figure}[!htb]
    \centering
    \includegraphics[clip,width=\linewidth]{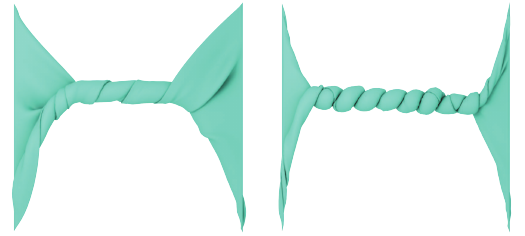}
    \caption{\figname{Mat-Twist} We reproduce the mat-twist example from~\cite{Li2020IPC}, demonstrating, overall, our method is similarly robust to the original \ac{IPC}. Left: our simulation at \qty{10}{\s} after 2 rounds of twisting at both ends. Right: at \qty{40}{\s} after 8 rounds of twisting. }
    \label{fig:mat-twist}
\end{figure}

Second, we reproduce the dolphin funnel (\cref{fig:dolphin}), trash compactor (\cref{fig:trash-compactor}), and mat twist (\cref{fig:mat-twist}) examples. Each of these examples features large deformations and complex contacts. Just as in \cite{Li2020IPC}, we robustly handle these scenarios and prevent intersections and inversions at every step. We note that in \cite{Li2020IPC} the fixed corotational model, which allows element inversion, was used instead of NeoHookean, so the number of iterations and timing reported in \cite{Li2020IPC} is significantly lower. To make fair comparisons, we also run \cite{Li2020IPC} with NeoHookean and report the statistics in Table~\ref{tab:params}. We refer to \cite{Flesh2018} for the artifacts in the fixed corotational model and comparisons with NeoHookean. For the dolphin funnel, we report the breakdown timing of our method: The Hessian assembly takes 34\%, linear solve 28\%, narrow phase CCD 19\%, broad phase CCD 4\%, and line search (excluding CCD) 14\%.

\begin{figure}[!htb]
    \centering
    \includegraphics[width=\linewidth]{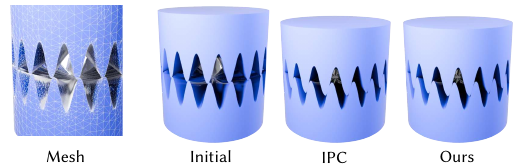}
    \caption{\figname{Monkey saddle} We force the top surface of the object on top to rotate in the counter-clockwise direction and fix the bottom surface of the object at the bottom. Our method is able to handle this scene with complex normal directions.}
    \label{fig:monkey_saddle}
\end{figure}

In \cref{fig:monkey_saddle}, as a stress test for our local minimum and exterior direction constraints, we generate an adaptive mesh for an n-legged monkey saddle, for which the normal oscillates around the center. For IPC, we use the maximum $\dhat$ that does not activate contact forces at the rest configuration, which is $5\times 10^{-5}$~\unit{\meter}. Since our method allows $\epsilon_\text{trg}$ larger than the edge length without activating contact, we make use of this advantage and set $\epsilon_{\text{trg}}=2\times 10^{-4}$~\unit{\meter}. Both methods produce similar results, our method takes 4255 iterations and 3.7 hours (1/3 less than IPC), while IPC takes 6710 iterations and 5.7 hours.

\begin{figure}[!htb]
    \centering
    \includegraphics[width=\linewidth]{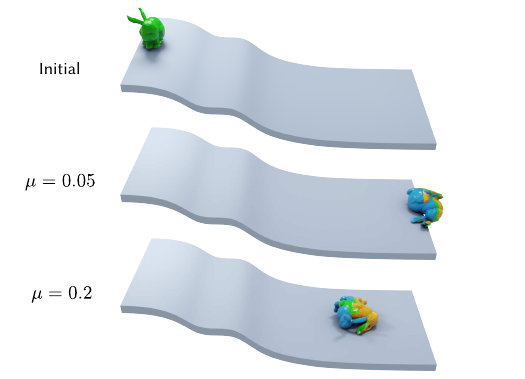}
    \caption{\textbf{Friction: Sliding bunny.} A bunny is thrown from the top of the slope. We simulate with friction coefficients 0.05 and 0.2, and with our method $\alpha=0.5$ (blue), Convergent IPC~\cite{Li2023Convergent} (yellow), and IPC~\cite{Li2020IPC} (green). The results at \qty{2}{\s} are shown.}
    \label{fig:bunny_friction}
\end{figure}

Finally, to validate our friction model, we simulate the bunny sliding on a slope with various friction coefficients, and with our method, Convergent IPC~\cite{Li2023Convergent}, and IPC~\cite{Li2020IPC} (\cref{fig:bunny_friction}). All three methods produce similar results.

\subsection{Inverse Design}\label{sec:inverse-design}

\begin{figure*}[!htb]
    \centering
    \includegraphics[width=\linewidth]{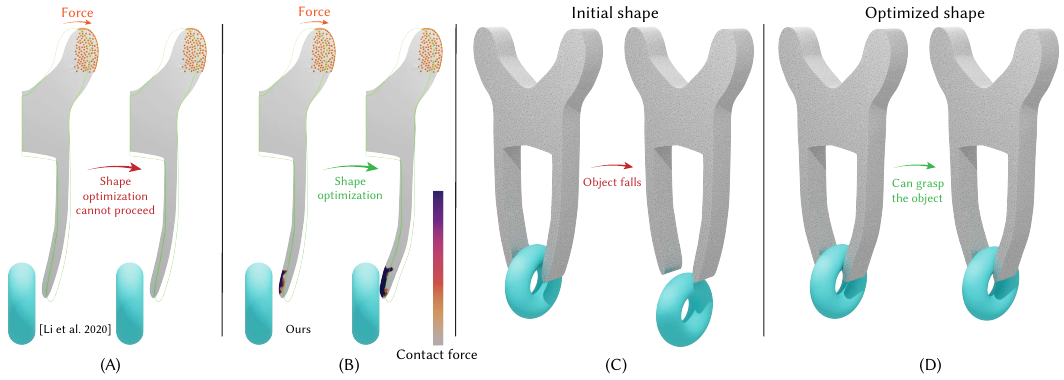}
    \caption{Shape optimization of a plier for grasping a torus. The handles (orange nodes) are pulled outwards (specified as the Dirichlet boundary conditions) in the simulations. Rest shapes and deformed shapes are visualized as green contours and grey surfaces respectively. (A) Shape optimization with \cite{Li2020IPC} cannot proceed, since there is no contact force between the torus and plier due to the restriction that $\dhat$ should be small to avoid spurious forces. (B) The contact force is maximized in the shape optimization with our method. (C) The initial shape fails to grasp the torus under gravity. (D) The optimized shape manages to grasp the torus under gravity. }
    \label{fig:inverse-design-plier}
\end{figure*}

We perform a shape optimization example to show one advantage of using a large $\dhat$ in the simulations. In \cref{fig:inverse-design-plier}, we optimize the shape of a plier so that it can grasp the torus under gravity when its handles are pulled outwards. We define the objective to be the contact force magnitude in the simulation and maximize the objective with L-BFGS. To reduce the dimension of the shape design space, we use the shape representation in \cite{Pneumatic2024} and sample 15 control points on the plier surface. We compute shape derivatives of the objective following \cite{DiffIPC2024}. For efficiency, we simulate only half of the plier and set symmetric boundary conditions. 

As a baseline, we first run the shape optimization with \cite{Li2020IPC} (\cref{fig:inverse-design-plier} A). Due to the restriction that $\dhat$ should be smaller than the minimum edge length of the mesh (to avoid spurious contact forces in Section \ref{sec:spurious-eval}), there is no contact force on the initial shape since the distance between the torus and the plier is larger than $\dhat$. Therefore, the shape derivatives of the objective are zero and the shape optimization cannot proceed. With our method (\cref{fig:inverse-design-plier} B), however, the $\dhat$ can be larger than the edge length without creating spurious forces between adjacent mesh vertices, so we can pick a large enough $\dhat$ so that there is nonzero contact force between the plier and the torus, and the shape optimization can proceed. The optimized plier manages to create enough contact forces to grasp the torus (\cref{fig:inverse-design-plier} D).

\subsection{Comparison to Repulsive Surfaces}\label{sec:repulsive}

\begin{figure}[!htb]
    \centering
    \includegraphics[width=\linewidth]{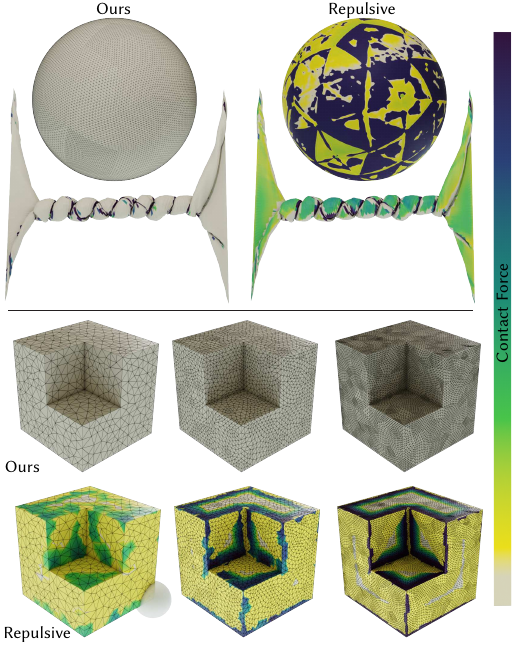}
    \caption{Potential gradient distributions with our formulation and the repulsive formulation for three examples: (A) Single sphere with radius \qty{1}{\m} and $\epsilon_\text{trg}=\qty{0.1}{\m}$, with magnitude range $[10^{-10}, 10]$. (B) Twist-mat with $\epsilon_\text{trg}=\qty{0.01}{\m}$, with magnitude range $[10^3, 10^{12}]$. (C) Cube corner with $\epsilon_\text{trg}$ visualized as a white sphere on the bottom left, with magnitude range $[10^{-1}, 10^5]$.}
    \label{fig:compare-repulsive}
\end{figure}

Tangent-point energy (TPE) of \citet{Yu2021RepulsiveSurfaces}, discussed in Section~\ref{sec:related-barrier}, has global support, which makes it expensive in simulations,
 even with acceleration.  To adapt it to our purposes, we multiply the integrand with a cubic spline, to localize it: 
\begin{equation}\label{eq:repulsive}
\iint_{M^2} \frac{\|P_f(x)(f(x) - f(y))\|^p}{\|f(x)-f(y)\|^{2p}} h_\epsilon(\|f(x)-f(y)\|) ~\mathrm{d}x_f\mathrm{d}y_f,
\end{equation}
and compare with our method in \cref{fig:compare-repulsive}. 

We observe that the TPE-based potential \cref{eq:repulsive} has a non-zero gradient everywhere on the surface if $\epsilon_\text{trg}$ is larger than the edge length. On a single sphere (\cref{fig:compare-repulsive}), the gradient of \cref{eq:repulsive} is small but nonzero everywhere, while ours is exactly zero. E.g., in the twist-mat example, both formulations have large forces at places in contact. However, the gradient of our potential vanishes at places that do not need the contact barrier, while the gradient of \cref{eq:repulsive} does not vanish anywhere.  

Furthermore, the TPE potentials are designed to optimize surfaces for smoothness: nonsmooth surfaces have infinite TPE \cite{Yu2021RepulsiveSurfaces}.  In our case, our goal is to be able to preserve sharp features of the original shape.

 We perform a convergence test for both formulations on a cube corner. Our potential is exactly zero for all 3 resolutions. For \cref{eq:repulsive}, forces concentrate around (both convex and concave) sharp corners and increase under mesh refinement. As expected (and desired for smooth surface optimization) \cref{eq:repulsive} converges to infinity under mesh refinement for sharp features), while our potential converges to a finite number when there is a (convex or concave) corner (\cref{fig:potential-converge}) and no contact.

\subsection{Parameter Dependence and Convergence}

To study the effects of our parameters we vary $\alpha$ and $\epsilon_\text{trg}$.

\begin{figure}[!htb]
\centering
\includegraphics[width=\linewidth]{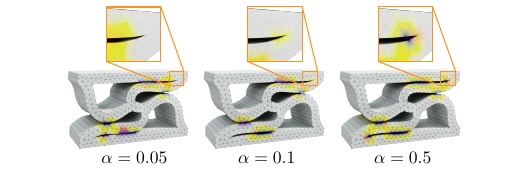}
\parbox[l]{0.2\linewidth}{Iterations:}
\parbox[c]{0.24\linewidth}{867}
\parbox[c]{0.24\linewidth}{625}
\parbox[c]{0.24\linewidth}{497}
\caption{\figname{Parameter-study: $\alpha$} Simulation of \cref{fig:unit-cell} with different $\alpha$ including the total number of solver iterations. As $\alpha$ increases, the nonlinear problem becomes softer, hence fewer iterations. However, large $\alpha$ introduces artifacts and spurious stresses in some cases.}
\label{fig:unit-cell-alpha}
\end{figure}

\begin{figure}[!htb]
    \centering
    \includegraphics[width=\linewidth]{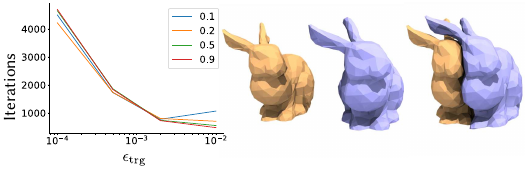}
    \caption{\figname{Parameter-study: $\epsilon_\text{trg}$} Simulation of two bunnies colliding with varying $\alpha$ and $\epsilon_\text{trg}$. We show the plot (left) of the number of iterations over $\epsilon_\text{trg}$, with a different $\alpha$ for each curve, as well as the initial frame (middle) and the colliding frame (right) of the simulation. For simple scenes, for a large range of $\alpha$, the number of iterations is similar; as $\epsilon_\text{trg}$ increases, it takes fewer iterations to converge.}
    \label{fig:parameter-study-dhat}
\end{figure}

Lower values of $\alpha$ (\cref{fig:unit-cell-alpha}) make the solution more accurate, as {\interaction} sets are further from points, and spurious forces do not appear even for extreme deformations, but the potential is less smooth and the solver requires more iterations. This effect is more pronounced in scenes with cusps; for simpler scenes, the difference in iteration count is less noticeable (\cref{fig:parameter-study-dhat}).

To study how much $\alpha$ helps to remove unnecessary contact pairs, we compute the number of contact pairs for varying $\alpha$ and $\epsilon_\text{trg}$ on the fixed scene in \cref{fig:mat-twist}, at the frame obtained after running the simulation for \qty{40}{\s} of simulation time, and compare against IPC \cite{Li2020IPC} and Convergent IPC \cite{Li2023Convergent} (\cref{tab:n_contact}). Convergent IPC has more contact pairs than IPC due to the extra contact pairs with negative weights introduced. Although our method has more types of contact pairs than IPC, even if the local minimum constraint is inactive ($\alpha=1$), it has fewer pairs than IPC due to the exterior direction constraint. The number of contact pairs drops significantly as $\alpha$ decreases, one can use reasonably large $\alpha$ for fast convergence while reducing the artifacts.

\begin{table}[!htb]
    \caption{\textbf{Number of contact pairs.} We compute the number of contact pairs in the fixed scene in \cref{fig:mat-twist} at \qty{40}{\s} with varying $\epsilon_\text{trg}$, using IPC \cite{Li2020IPC}, Convergent IPC \cite{Li2023Convergent}, and our method with different choices of $\alpha$.}
    \footnotesize
    \centering
    \resizebox{\linewidth}{!}{\begin{tabular}{c|ccccccc}
        $\epsilon_\text{trg}$ & IPC & Convergent & $1$ & $0.8$ & $0.5$ & $0.1$ \\\hline
        $0.001$ & 226k & 250k & 215k & 176k & 128k & 53k \\
        $0.002$ & 372k & 405k & 322k & 225k & 155k & 56k \\
        $0.005$ & 1690k & 1772k & 830k & 433k & 273k & 78k \\
    \end{tabular}}
    \label{tab:n_contact}
\end{table}

We plot in \cref{fig:parameter-study-dhat} the effect of $\epsilon_\text{trg}$ on the number of iterations. As the support of the contact potential increases the problem becomes softer and the number of iterations decreases accordingly. 

\begin{figure}
    \centering
    \includegraphics[width=\linewidth]{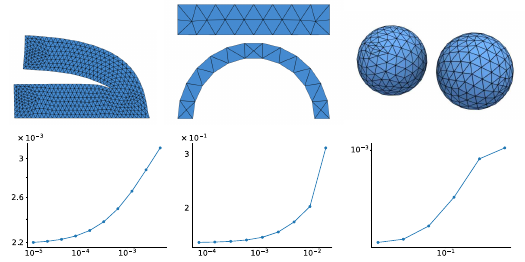}
    \caption{Convergence of potential under mesh refinement in fixed scenes with fixed $\epsilon_\text{trg}$. We pick 2 configurations in 2D and 1 in 3D, start with the coarse mesh, and compute the potential under mesh refinement. The plots of contact barrier potential over edge length are shown.}
    \label{fig:potential-converge}
\end{figure}

We also perform a convergence study for three different scenarios shown in \cref{fig:potential-converge}. We see convergence under mesh refinement in all of these scenes. Importantly, we use a fixed $\epsilon_\text{trg}$ (unlike Convergent IPC~\cite{Li2023Convergent} which requires co-refinement of $\dhat$).

We note that when our discretization is applied to the piecewise linear surface, treating it as a piecewise smooth surface, the integrals for edge potentials and summations for vertex potentials are introduced, with the relative scale of potentials determined by the constant $L$.  As the surface is refined, the scale of the constant also needs to be adjusted, for the refined mesh potential to approximate the smooth surface potential. We leave a rigorous study of convergence of the discrete potential to the potential of the limit smooth surface as future work. 

\subsection{Infinite Potential}\label{sec:infinite-potential}

\begin{figure}
    \centering
    \includegraphics[width=0.5\linewidth]{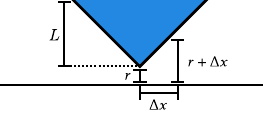}
    \caption{Setup for corner hitting a plane discussed in \cref{sec:infinite-potential}.}
    \label{fig:infinite-potential-setup}
\end{figure}

\begin{figure*}
    \centering
    \includegraphics[width=\linewidth]{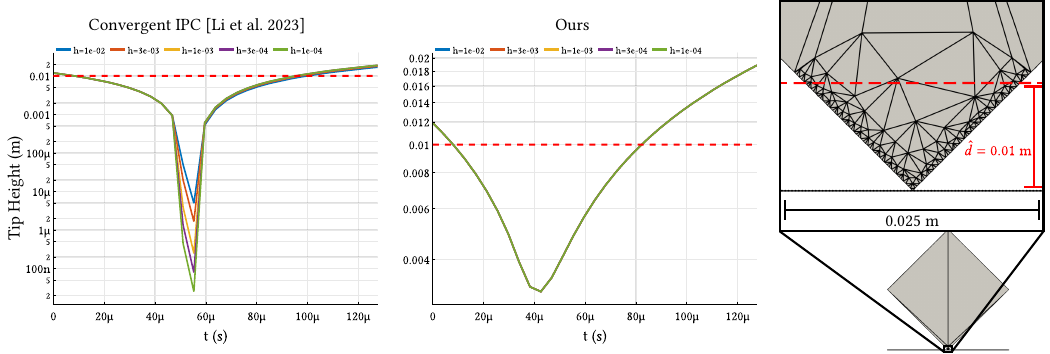}
    \caption{\figname{Finite Potential for Zero Distance} A square impacts a plane at the corner of the square with a velocity of \qty{234}{\m\per\s}. We refine both meshes (only at the corner of the square for efficiency). Here we plot the height of the squares tip over time. Left: for Convergent IPC~\cite{Li2023Convergent}, we see that the minimum distance shrinks with every refinement of the mesh. This is a consequence of the continuous model having a finite potential. Middle: due to our choice of barrier function, our method exhibits the same trajectory (all plots overlap) under refinement.}
    \label{fig:corner-hit}
\end{figure*}

Consider a corner hitting a plane, and suppose they are both refined (\cref{fig:corner-hit} Right). In this case, the integral of the Convergent IPC~\cite{Li2023Convergent} potential over the plane will be finite, even if the apex of the corner is directly on the plane (violation of \cref{req:barrier}). Although each discretization will be infinite, the implication is that the total potential will decrease as we refine so that the actual distance will decrease to zero. 

\Cref{fig:infinite-potential-setup} shows the configuration: for a gap of $r$ and a distance from the closest point on the plane $\Delta x$, points along the square will be at a distance of $\Delta x + r$. Assume the potential\footnote{The extra quadratic term does not affect the conclusions.} for points on the corner is just $\log\left(\frac{\Delta x + r}{\dhat}\right)$, then integrating over $x$ we get 
\[
\int_0^L \log\left(\frac{x+r}{\dhat}\right) dx = \left( r+L \right) \log  \left( {\frac {r+L}{\dhat}} \right) -\log  \left( {\frac {r}{\dhat}} \right) r-L,\] 
and for $r \rightarrow 0$ (i.e. approaching zero distance), it has a finite limit of $\log  \left( {\frac {L}{\dhat}} \right) L-L$. 

The force (i.e., the potential derivative) at $r = 0$ is $\sim\log(r)$ though (i.e. is infinite), so for a finite force the static equilibrium problem will always have a valid solution. This is not the case however for a dynamic problem: if the kinetic energy exceeds the finite potential, then the barrier cannot prevent contact. We show this case in \cref{fig:corner-hit}. One can see the Convergent IPC model results in ever decreasing distances as the mesh is refined (without refining $\dhat$). Our method in comparison exhibits the same trajectory for all levels of refinement. This is because we use a barrier function whose integral over the surface is not finite as the distance goes to zero and because the potential converges under mesh refinement.

Given that the minimum distance does not depend on the mesh resolution, one can pick $\epsilon_\text{trg}$,  independent of the edge lengths, based on the desired accuracy of the contact handling.

\section{Conclusion}

Barrier potentials have enabled a qualitative improvement in the robustness of accurate simulation of deformable objects with complex contact. In this paper, we revisit barrier-based contact formulations and relate them to a family of potentials defined for a broad class of surfaces that satisfy a set of natural requirements. We demonstrate that applying these principles leads to a new formulation that alleviates some of the shortcomings of existing methods for barrier-based contact simulation. 

\subsection{Future work} Our derivation assumes closed surfaces without boundaries at several steps, which needs modifications for codimensional objects; a possible direction for future work is handling surfaces with boundaries, and codimensional surfaces. Besides, as \citet{Du2023Free} pointed out, IPC~\cite{Li2020IPC} produces spurious tangential contact forces on a flat plane with non-uniform discretization, thus, it cannot simulate well the free-sliding of a cube on a plane without friction. Our method does not resolve this artifact. A potential solution is to use a high-order quadrature to better integrate the continuous formulation (in addition to including the closest-point quadrature), so that the potential is less affected by the non-uniform discretization. Another direction is deriving a higher-order discretization of our continuum formulation to increase accuracy.

\begin{table*}[!htb]
    \centering
    \caption{\figname{Simulation statistics} For each simulation we report geometry, (minimum, average, and maximum edge length $h$), time step $\Delta t$, material (Young's modulus $E$, Poisson ratio $\nu$, and density $\rho$), $\epsilon_\text{trg}$, maximum memory used, as well as average timing and number of Newton iterations. A value of 0 for iterations indicates the optimization did not run because the initial configuration was at a force equilibrium (i.e., no spurious forces at rest). Timings and iterations of \cite{Li2020IPC} are shown in parentheses.}
    \footnotesize
    \begin{tabular}{l|c;{1pt/1pt}c;{1pt/1pt}c;{1pt/1pt}c;{1pt/1pt}c;{1pt/1pt}c;{1pt/1pt}c;{1pt/1pt}c}
    Example & \# nodes, \# cells & \makecell{h (\unit{\m})\\(min, avg, max)} & $\Delta t$ (\unit{\s}) & \makecell{\(E\) (\unit{\kilo\Pa}), \(\nu\),\\\(\rho\) (\unit{\kg\per\m\cubed})} & \(\epsilon_\text{trg}\) (\unit{\m}) & $\alpha$ & \makecell{memory\\(\unit{\mega\byte})} & \makecell{timing (\unit{\s}),\\iterations\\(per timestep)}\\
    \hline
      Slit block (2D) (Supp. Video) & 199, 325 &  0.04, 0.18, 0.34 &  0.0047 &   4, 0.2, 100 &  0.1 &    0.1 & 19 & 0.07, 0\\
       Slit block (3D) (Fig~\ref{fig:spurious-stresses-rest-3D}) & 1058, 3871 &  0.01, 0.04, 0.13 &  0.0047 &    0.1, 0.2,  10 &  0.0125 &    0.1 & 93 & 0.14, 0\\
    Fillet block (2D) (Fig~\ref{fig:spurious-stresses-nonuniform-2D}) & 445, 672 &  0.01, 0.04, 0.21 &  0.0047 &   4, 0.2, 100 &  0.1 &    0.1 & 26 & 0.09, 0 \\
     Fillet block (3D) (Fig~\ref{fig:teaser}) & 1003, 3824 &  0.02, 0.19, 0.42 &  0.0047 &    0.1, 0.2,  10 &  0.07 &    0.1 & 92 & 0.14, 0\\
Compressed block (2D) (Supp. Video) & 356, 630 &  0.09, 0.12, 0.22 &  0.0047 &  10, 0, 100 &  0.04 &    0.1 & 63 & 0.17, 2.3 \\
Compressed block (3D) (Fig~\ref{fig:spurious-stresses-compressed}) & 2169, 9799 &  0.03, 0.10, 0.18 &  0.0047 &  10, 0, 100 &  0.0125 &    0.1 & 644 & 0.78, 2.5 \\
        Armadillo bar (\cref{fig:armadillo-sim}) & 91777, 446158 & 0.00076, 0.0090, 0.084 & 0.02 & 10, 0.49, 1 & 0.001 & 0.5 & 9675 & 22.2, 3.4 (64.4, 10.4)\\
      Two blocks (2D) (Supp. Video) & 708, 1260 &  0.09, 0.12, 0.22 &  0.0047 &   4, 0.2, 100 &  0.1 &    0.1 & 24 & 0.08, 0\\
      Two blocks (3D) (Fig~\ref{fig:spurious-stresses-rest-3D}) & 580, 2079 &  0.04, 0.21, 0.55 &  0.0047 &  10, 0, 100 &  0.025 &    0.1 & 61 & 0.14, 0\\
      Spherical balloon (3D) (Fig~\ref{fig:balloon}) & 421, 701 &  0.04, 0.11, 0.18 &  0.0047 &   4000, 0, 100 &  0.02 &    0.1 & 722 & 0.67, 1\\
         Donut balloon (3D) (Fig~\ref{fig:balloon-donut}) & 17122, 52601 &  0.013, 0.3, 0.086 &  0.01 &  1000, 0.48, 1000 &  0.01 &    0.1 & 2205 & 28.9, 17.1\\
       Cusp compression (Fig~\ref{fig:unit-cell}) & 3761, 12777 & 0.024, 0.047, 0.089 & 0.01 & 1000, 0.3, 100 & 0.015 & 0.05 & 501 & 5.02, 15.8\\
       Dolphin funnel (Fig~\ref{fig:dolphin}) & 4074,10511 & 0.0017,0.020,0.081 & 0.025 & 10,0.4,1000 & 0.001  & 1 &  1403 & 15.4, 58.1 \\
        Trash compactor (Fig~\ref{fig:trash-compactor}) & 6611, 21696 & 0.00033,0.049,0.36 & 0.01 & 10, 0.4, 1000 & 0.001 & 0.8 & 2833 & 228.9, 155 (196, 205) \\
        Mat twist (Fig~\ref{fig:mat-twist}) & 45000, 133206 &0.0067, 0.0088, 0.0126  & 0.04 & 20,0.4,1000  & 0.002  & 0.8 & 17683 &  697.7, 61.8 (610.3, 116) \\
        Monkey saddle (\cref{fig:monkey_saddle}) & 67320, 198282 & 0.00032, 0.0080, 0.058 & 0.005 & 1e5, 0.48, 1000 & 0.0002 & 0.8 & 5376 & 66.6, 21.3 (102.6, 33.6)
\end{tabular}

    \label{tab:params}
\end{table*}

\begin{acks}
This work was supported in part through the NYU IT High Performance Computing resources, services, and staff expertise. This work was also partially supported by the NSF CAREER award under Grant No. 1652515, the NSF grants OAC-2411349, OAC-1835712, CHS-1908767, CHS-1901091, IIS-2313156, a Sloan Fellowship, and a gift from
Adobe Research.

This material is based upon work supported by the U.S. Department of Energy, Office of Science, Office of Advanced Scientific Computing Research, Department of Energy Computational Science Graduate Fellowship under Award Number(s) DE-SC0025528.

This report was prepared as an account of work sponsored by an agency of the United States Government. Neither the United States Government nor any agency thereof, nor any of their employees, makes any warranty, express or implied, or assumes any legal liability or responsibility for the accuracy, completeness, or usefulness of any information, apparatus, product, or process disclosed, or represents that its use would not infringe privately owned rights. Reference herein to any specific commercial product, process, or service by trade name, trademark, manufacturer, or otherwise does not necessarily constitute or imply its endorsement, recommendation, or favoring by the United States Government or any agency thereof. The views and opinions of authors expressed herein do not necessarily state or reflect those of the United States Government or any agency thereof.
\end{acks}

\bibliographystyle{ACM-Reference-Format}
\bibliography{99-bib}

\appendix

\section{Distance function mollification}
\label{sec:distance-mollification}
\begin{figure}
    \centering
    \includegraphics[width=\linewidth]{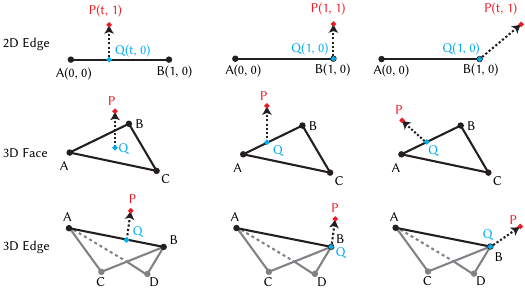}
    \caption{Non-smoothness of the closest point position. From left to right, the closest point $Q$ of a point $P$ outside of an edge/face moves from the interior of the edge/face to its boundary. The coordinate of $Q$ does not depend smoothly on $P$ as $P$ crosses the boundary.}
    \label{fig:nonsmooth-closest-point}
\end{figure}
When differentiating the potential, since the closest points are used for evaluating the quadrature, one needs to differentiate through the closest points with respect to the vertex positions of the mesh. However, the closest points may not smoothly depend on the vertex positions. For example, in 2D (top row in \cref{fig:nonsmooth-closest-point}), the closest point on the unit edge $AB$ to a point $P=(t, 1)\ (t>0)$ is $Q=(\min(t,1), 0)$, which is only $C^0$ continuous at $t=1$. Below we introduce new types of mollification different from \cite{Li2020IPC} to resolve this problem. First, we define $d(\cdot, \cdot)$ as the shortest distance between two primitives (vertex, edge, or face), e.g. $d(P, A)$ denotes the distance between two points $P$ and $A$, $d(P, AB)$ %
denotes the distance of point $P$ to edge $AB$. We reuse the $C^1$ continuous mollifier function from \cite{Li2020IPC}:
$$
h(z):=\begin{cases}
z(2-z) & 0 \leq z < 1 \\
1 & 1\leq z \\
\end{cases}\quad
h_c(s):= h(\frac{s-1}{c}),
$$
we use $c=0.01$ in our examples. In the definition of $M(x,y)$, we denote $P=f(x)$ and $Q=f(y)$ as the pair of closest points.

For Edge-Vertex contact (either in 2D or 3D), suppose $Q$ is on an edge $AB$, we define the mollification as 
$$
M(x,y):=h_c( \frac{ d(A, P) }{ d(P, AB) }) h_c( \frac{ d(B, P) }{ d(P, AB) }),
$$
which vanishes as the closest point $Q$ approaches either $A$ or $B$ (\cref{fig:nonsmooth-closest-point}). For Face-Vertex contact, let the vertices of the triangle face be $A, B, C$, and the vertex outside be $P$ (\cref{fig:nonsmooth-closest-point}). Similarly, the mollification is 
$$
M(x,y):=h_c( \frac{ d(P, AB) }{ d(P, ABC) }) h_c( \frac{ d(P, AC) }{ d(P, ABC) }) h_c( \frac{ d(P, BC) }{ d(P, ABC) }),
$$
which vanishes as $Q$ on the triangle $ABC$ approaches the boundary of the triangle.

\begin{figure}
    \centering
    \includegraphics[width=\linewidth]{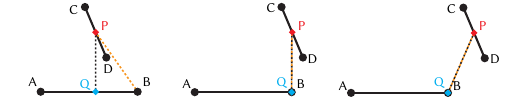}
    \caption{Non-smoothness of the Edge-Edge contact. Point $Q$ and $P$ are the closest point pair between edge $AB$ and $CD$, the orange line is the closest direction between vertex $B$ and edge $CD$. As edge $CD$ moves from left to right, $Q$ is only $C^0$ continuous when it reaches $B$. At the same time, $BP$ starts to overlap with $QP$.}
    \label{fig:mollifier-edge-edge}
\end{figure}

For Edge-Edge contact, the smoothness is more complicated since $P$ is not smooth anymore. In \cref{fig:mollifier-edge-edge}, we observe that the closest points $P,\ Q$ are non-smooth only when the Edge-Edge distance reduces to Vertex-Edge or Vertex-Vertex distance. E.g. in \cref{fig:mollifier-edge-edge}, line $BP$ overlaps with $QP$ when $Q$ is only $C^0$ continuous. Inspired by this, we define our mollification as 
$$
\resizebox{.95\hsize}{!}{$M(x,y):=h_c(\frac{d(B, CD)}{d(AB,CD)})h_c(\frac{d(A, CD)}{d(AB,CD)})h_c(\frac{d(C, AB)}{d(AB,CD)})h_c(\frac{d(D, AB)}{d(AB,CD)}),$}
$$
where $d(PQ,UV)$ is the distance between two edges $PQ$ and $UV$. The mollification is $C^1$ smooth and vanishes when $Q$ overlaps with $A$ or $B$. Note that the parallel edge-edge mollification in \cite{Li2020IPC}, which was introduced to avoid non-smooth distance with almost parallel edges, is not needed anymore, since our mollification also vanishes when the edges are in parallel.

All the distance functions used in the mollifications above are $C^\infty$ smooth as long as there is no intersection or degenerate element, and $h_c$ is $C^1$ smooth, so the mollifications are $C^1$ smooth (same as \cite{Li2020IPC}). 

\begin{remark}
The proposed mollifications make the contact potential vanish in more cases than in \cite{Li2020IPC}, however, it still guarantees intersection-free, i.e. the \cref{req:barrier} (\Rbar) is still satisfied: When an Edge-Vertex contact vanishes due to mollification, the closest point on the edge is at one of the endpoints and the Vertex-Vertex contact becomes active; when a Face-Vertex (or Edge-Edge) contact vanishes, some Edge-Vertex or Vertex-Vertex contact is active.
\end{remark}

\section{Determining Orientation in Piecewise Linear Case } 
\label{app:outside-filtering}

In this appendix, we detail our method for determining if $v:=(f(y) - f(x))_+$ at a point $y$ of an oriented piecewise linear surface points inside. For high-order surfaces, one only needs to consider local tangent planes at $y$, so the problem reduces to the piecewise linear case.

\begin{figure}
    \centering
    \includegraphics[width=\linewidth]{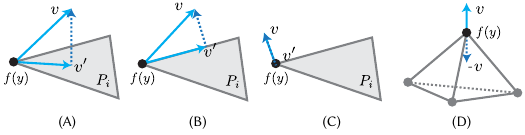}
    \caption{(A) The closest point $v'$ on face $P_i$ is in the interior of $P_i$. (B) The closest point $v'$ is on the boundary of $P_i$. (C) The closest point $v'$ overlaps with $f(y)$. (D) The closest points on all faces overlap with $f(y)$. In this case, instead of deciding if $v$ points inside, we flip $v$ and decide if $-v$ points outside.}
    \label{fig:cone-filter-algo-1}
\end{figure}

As shown in \cref{fig:normal-notation}, when contact happens (i.e. as $f(x)$ approaches $f(y)$), if the cone around $f(y)$ is convex, one may conclude that $n_i(y)\cdot v\leq 0$ for all $i$. However, in general, $n_i(y)\cdot v\leq 0$ only holds for some $i$, and $\min_i n_i(y)\cdot v<0$ may be extremely close to zero. As a consequence, even though one can simply drop contact pairs based on normals for \cite{Li2020IPC} in some simple scenes (\cref{fig:balloon}), in general, the filtering on normal directions has to be constructed to be both numerically robust and differentiable.

We first discuss the robust and exact algorithm (not differentiable). For convenience, let $f(y)$ be the origin of the coordinate system. let $e_i\ (i=1,\dots,M)$ be the vector pointing from vertex $f(y)$ to its $i$-th neighboring vertex, $P_i$ be the face formed by point $f(y)$ and vectors $e_i$ and $e_{i-1}$ (let $e_0=e_M$). Denote $n_i$ as the normal of $P_i$, the $\{e_i\}$ follow the ordering so that $(e_{i-1}\times e_i)\cdot n_i > 0$. We assume that the length of $e_i$ is much larger than $v$ since we only care about the direction of $v$. In the trivial case, where every $P_i$ is perpendicular to $v$, i.e. all $P_i$ share the same plane, we can define
$$
\Phi^e(x,y):=-v\cdot n,
$$
where $n$ is the normal shared by all $P_i$. Below we assume that there exists $P_i$ that is not perpendicular to $v$.

We first find the closest point to $v$ on the cone surface. In \cref{fig:cone-filter-algo-1}, we find the closest point to $v$ on each face $P_i$, and denote the projected vector as $v'_i$. Suppose 
$$
j=\text{argmin}_i \|v - v'_i\|,
$$
which means $v'_j$ is the closest point to $v$ over the whole surface of the cone formed by $f(y)$ and its 1-ring. As a consequence, the line segment connecting $v'_j$ and $v$ doesn't intersect with any other faces of the cone, i.e. the segment lies completely inside or outside of the cone. 
\begin{itemize}
    \item If $v'_j$ is on the interior of face $P_j$, we can define
        $$
        \Phi^e(x,y):=-v\cdot n_j,
        $$
        where $n_j$ is the normal of $P_j$.
    \item If $v'_j$ is shared by two faces, i.e. it lies on an edge, the problem reduces to 2D (\cref{fig:closest-shared}): we project the space onto the plane perpendicular to the edge $e_j$, and the problem becomes checking if 2D vector $\tilde{v}$ is inside the sector bounded by $\tilde{e}_1, \tilde{e}_2$. Since $\tilde{e}_1\times \tilde{n}_1 > 0$ (by construction of face normals), $\tilde{v}$ points inside if and only if $$\tilde{e}_1\rightarrow\tilde{e}_2\rightarrow\tilde{v}$$ is counter-clockwise. If $\tilde{v},\tilde{e}_1,\tilde{e}_2$ are all unit-length, then it's equivalent to 
    $$\Phi^e(x,y):=(\tilde{v}-\tilde{e}_1)\times (\tilde{v}-\tilde{e}_2) > 0.$$
    \item If $v'_j=f(y)$, i.e. the closest point to $v$ is the center of the cone. In this case, the closest point to $-v$ on the cone surface must not be $f(y)$, otherwise all faces $P_i$ are perpendicular to $v$ and it becomes the trivial case we discussed above. Thus we can perform the above algorithm on $-v$ instead: Deciding if $v$ points inside is equivalent to deciding if $-v$ points outside.
\end{itemize}

\begin{remark}
If the closest point $v'_i$ to $v$ on some face $P_i$ is not at $f(y)$, then $\|v-v_i'\| < \|v - f(y)\|$, because $f(y)$ is on all faces and the distance (which is a strictly convex function) reaches the minimum at $v'_i$ on face $P_i$.
\end{remark}

The $\Phi^e(x,y)$ defined above satisfies that $\Phi^e(x,y) > 0$ if and only if $v$ points inside the cone, and $\Phi^e(x,y) = 0$ if and only if $v$ lies on the surface of the cone.

We summarize the algorithm in \cref{algo:filter-cone} and describe the purpose of each function below.
\begin{itemize}
    \item \textsc{DetermineInside3D}$(v, \{e_i\}, \{n_i\})$ returns a flag indicating if $v$ points inside the cone bounded by rays $\{e_i\}$. $n_i$ is the normal of face bounded by rays $e_{i-1},e_i$.
    \item \textsc{ClosestPoint}$(v,\{e_i\}$ finds the closest point to $v$ on the cone surface.
    \item \textsc{FaceClosestPoint}$(v,a,b)$ returns the closest point to $v$ on the angle sector spanned by $a, b$, and a flag indicating whether the closest point is on the boundary or in the interior of the sector.
    \item \textsc{Project}$(a,b)$ projects vector $a$ onto the plane perpendicular to vector $b$, and normalize the projected vector.
    \item \textsc{EdgeClosestDistance}$(v,e)$ returns the distance of $v$ to the ray $e$.
\end{itemize}

\begin{algorithm}
\begin{algorithmic}
    \Function{DetermineInside3D}{$v, \{e_i\}, \{n_i\}$}
        \State $[v', \text{type}, j] \gets \Call{ClosestPoint}{v, \{e_i\}}$
        \If{$v'==\vec{0}$} \Comment{The closest point is $f(y)$}
            \If{$\{n_i\}$ are close to the same vector}
                \State \Return $-v\cdot n_0 > 0$
            \Else
                \State \Return \Call{DetermineInside3D}{$-v, \{e_i\}, \{n_i\}$} == False
            \EndIf
        \EndIf
        \If{$\text{type}$==Interior}  \Comment{The closest point is inside a face}
            \State \Return $-v\cdot n_j > 0$ 
        \Else \Comment{The closest point is on an edge}
            \For{$i=1$ to $M$}
                \State $d_i \gets \text{EdgeClosestDistance}(v, e_i)$
            \EndFor
            \State $j=\text{argmin}_i d_i$
            \State $\tilde{e}_1\gets \Call{Project}{e_{j-1}, e_j}$
            \State $\tilde{e}_2\gets \Call{Project}{e_{j+1}, e_j}$
            \State $\tilde{v}\gets \Call{Project}{v, e_j}$
            \State \Return $(\tilde{v}-\tilde{e}_1)\times (\tilde{v}-\tilde{e}_2) \cdot e_j < 0$
        \EndIf
    \EndFunction
    
    \Function{ClosestPoint}{$v, \{e_i\}$}
        \For{$i=1$ to $M$}
            \State $[p_i, \text{type}_i]\gets \Call{FaceClosestPoint}{v, e_{i-1}, e_i}$
        \EndFor
        \State $j\gets \text{argmin}_i \|p_i-v\|$
        \State \Return $[p_j, \text{type}_j, j]$
    \EndFunction
    
    \Function{EdgeClosestDistance}{$v, e$}
        \State $w\gets \max\{(v\cdot e)/\|e\|^2, 0\}$
        \State \Return $\|v-w e\|$
    \EndFunction
    
    \Function{Project}{$a, b$}
        \State $a'\gets a - (a\cdot b / \|b\|^2) b$
        \State \Return $a' / \|a'\|$
    \EndFunction
\end{algorithmic}
\caption{Determine if a vector $v$ is inside a cone bounded by a set of rays $\{e_i\}$.}
\label{algo:filter-cone}
\end{algorithm}

We note that if one uses the algorithm above to filter the contact pairs directly, the resulting contact potential is not continuous: In \cref{fig:cone-filter-setup}, if the vector $v$ lies on the intersection of the two sub-domains, under small perturbations, the contact pair may be dropped or kept in the filtering. Thus, the potential requires complicated mollification to guarantee $C^1$ smoothness with this algorithm. Instead, we use a simplified smooth formulation as in \cref{eq:normal_mollified}, which is not exact but conservative.

\section{Formulas for smoothing functions}
\label{app:smoothing}
\[
B^3(v) = \begin{cases}
{\frac{2}{3}}-{{v}^{2}+\frac {1}{2}|v|^{3}} & |v|<1\\
{\frac {1}{6} \left( 2-|v| \right)^3} & 1 \leq |v|<2 \\
0 & 2 \leq |v|. \\
\end{cases}
\] 

 \[
 H(z) = \begin{cases}
 0 &  z < -3\\
 \frac{1}{6}(3+z)^3 & -3 \leq z<-2\\
 \frac{1}{6}(3-9z-9z^2-2z^3) & -2 \leq z<-1\\
 1+\frac{1}{6}z^3 & -1 \leq z<0\\
 1 & 0 \leq z.
 \end{cases}
 \]

\section{Details of the p.w. smooth formulation}
\label{sec:pw-smooth-appendix}

\subsection{Details of the local minimum constraint}
If $y$ is on more than one patch, i.e. on the shared boundary of patches, on each patch $\boundary_i$, there are two tangent vectors $t^1_i(y)$ and $t^2_i(y)$, along two edge curves of $\boundary_i$ meeting at $y$ (\cref{fig:tangent-notation}). If $\transpose{t^j_i(y)}~(f(y)-f(x))_+\geq 0$, then for any $t=\sum_{j=1}^2 a_j t^j_i(y)~(a_j\geq 0)$, we have
$$
\transpose{t}~(f(y)-f(x))_+ = \sum_{j=1}^2 a_j \transpose{t^j_i(y)}~(f(y)-f(x))_+\geq 0,
$$
which means \cref{eq:parametric_minima} holds for any $t$ in the convex cone spanned by $t^1_i(y)$ and $t^2_i(y)$ (\cref{fig:tangent-notation} A). However, the tangent directions of $\Omega_i$ at $y$ may not always lie in the cone. For example, when $y$ is on an edge of a linear triangle as $\boundary_i$, $t^1_i(y) + t^2_i(y) = 0$, then the convex cone degenerates to a line, while the tangent directions form a half-plane (\cref{fig:tangent-notation} B); when $y$ is at a concave corner of a high order patch $\boundary_i$, then the convex cone is the complement of $\boundary_i$ (\cref{fig:tangent-notation} C).

To treat all cases uniformly, we add a halfway vector $t^3_i$, corresponding to the angle bisector of the angle between $t^1_i$ and $t^2_i$ in $\boundary_i$ (\cref{fig:tangent-notation}). Then every tangent vector $t$ of $\Omega_i$ at $y$ can be represented as a convex combination of $\{t^j_i\}_{j=1}^3$, and consequently, \cref{eq:parametric_minima} along these three directions ensures that the point is a local minimum of the distance.
 
\subsection{Local minimum term} The mollification of local minimum constraints $\Phi^m_{ik}(x,y)\geq -\alpha,\ \forall i\in I$ is straightforward and similar to the smooth surface case:
$$
g^m(x,y):=\prod_{i\in I}\prod_{k=1}^{3} H^{\alpha}(\Phi^m_{ik}(x,y))
$$
reaches maximum of 1 when $\Phi^m_{ik}(x,y)\geq 0,\ \forall i\in I$, and vanishes if there exists $i\in I$ such that $\Phi^m_{ik}(x,y)\leq -\alpha$.

Since $\Phi^m_{ik}$ is a dot product between unit vectors, $\alpha$ acts as a threshold of contact potential on the cosine value of the angle between contact pairs. One can pick $\alpha$ based on the desired angle threshold, we use $0.05\leq \alpha\leq 0.8$ in the examples. Although extremely small $\alpha$ leads to numerical issues in the simulation, \cref{req:barrier} is satisfied as long as $\alpha>0$, i.e., the simulation is always contact-free.

\subsection{Exterior direction term} For edge points in Edge-Edge and Edge-Vertex contact, based on \cref{eq:cone-filter-2D}, we can define 
$$
g^e(x,y) := H^\beta (\Phi^e(x,y)),
$$
where $\beta$ is a parameter that controls the smoothness of exterior direction constraints. $g^e(x,y)$ reaches maximum when $\Phi^e(x,y)\geq 0$ and vanishes if $\Phi^e(x,y)\leq -\beta$. 

For vertices in Face-Vertex, Edge-Vertex, and Vertex-Vertex contact, we first define 
$$
\Phi^e_{i}(x,y):=-\transpose{n_i(y)} (f(y)-f(x))_+.
$$
However, unlike the case of the smooth surfaces, there are multiple normal directions at one vertex (\cref{fig:normal-notation}). To eliminate the contact between opposite sides of the object (Figure~\ref{fig:candidates}), the following should hold for some $i\in I$
\begin{equation}\label{eq:normal}
    \Phi^e_{i}(x,y) \geq 0 .
\end{equation}
To make the constraint smooth, we relax this to a simpler, weaker condition. We define 
\begin{equation}\label{eq:normal_mollified}
g^e(x,y) :=H^{1}\left(-1+\sum_{i\in I} H^{\beta}\left(\Phi^e_{i}(x,y)\right)\right),
\end{equation}
which reaches a maximum of 1 when there exists $i$ such that
$
\Phi_{i}^e(x,y) \geq 0,
$
and vanishes if for all $i$,
$
\Phi_i^e(x,y) \leq -\beta.
$

\section{Exclusion of adjacent elements}
\label{sec:adjacent-exclusion}

Adjacent elements such that one does not contain another can be vertex-adjacent face and edge, and vertex- and face-adjacent faces.   Suppose an edge and a face are in contact, i.e., an edge is contained in the plane of the face, and they share a vertex and other points. Then the other edge endpoint is either inside the face, or the edge intersects the boundary of the face, crossing an edge which it does not share a vertex with.  In the first case, we have vertex-face contact, in the second case, we have non-adjacent edge-edge contact. Thus, even if we do not apply the potential to the adjacent edge-face pair, when they are close to contact, it will be activated for a non-adjacent element pair.  If it is a face-face contact, the argument reduces to the first, if applied to one of the face edges that is not a common edge with the other face.

\clearpage

\end{document}